\documentclass{aastex63}
\usepackage{graphicx}
\usepackage{float}
\usepackage{natbib}
\usepackage{longtable}
\usepackage{hyperref}
\usepackage{scrextend}
\usepackage{listings}
\usepackage{xcolor}

\begin{document}
\title{A Tale of Three Dust Populations: Variable $R_{\rm{V}}$ and Extreme Polarization Along Sightlines Toward $\zeta$~Ophiuchi}
\author{Ashley N. Piccone}
\affil{University of Wyoming \\
1000 E University Ave \\
Laramie, WY 82071}
\author{Henry A. Kobulnicky}
\affil{University of Wyoming \\
1000 E University Ave \\
Laramie, WY 82071}
\correspondingauthor{Ashley Piccone}
\email{apiccone@uwyo.edu}

\begin{abstract}
        Dust permeates the interstellar medium, reddening and polarizing background starlight,  but dust properties vary with local environment. In order to characterize dust in a highly irradiated diffuse cloud, we measure the reddening and optical polarization towards 27 stars surrounding the mid-latitude $b$=$+$24\degr ~O9.2IV star $\zeta$~Ophiuchi, using new optical spectroscopy and polarimetry. We incrementally deredden and depolarize with distance, allowing us to distinguish dust components along these sightlines. The data indicate three distinct dust populations: a foreground component characteristic of average Milky Way dust ($R_{\rm{V}}$$\approx$3.1, $d$$\lesssim$180~pc), a highly polarizing mid-distance component in the vicinity of $\zeta$~Oph ($R_{\rm{V}}$$\approx$2.4, 200~pc$<$$d$$<$300~pc), and a non-polarizing distant component ($R_{\rm{V}}$$\approx$3.6, 600~pc$<$$d$$<$2000~pc). Prominent 8~$\mu$m infrared striations spanning the field of view likely have high Polycyclic Aromatic Hydrocarbon content and are illuminated by $\zeta$~Oph. Foreground-subtracted polarizations roughly align with these striations, which, we argue, lie immediately behind $\zeta$~Oph and constitute the highly-polarizing mid-distance dust. This component polarizes very efficiently ($P_{\rm{V}}$$>$9.1$E(B-V)$), implying a high degree of grain alignment and suggesting that the bulk of the polarization occurs in a small fraction of the volume. The large $R_{\rm{V}}$ in the distant component reveals that dust above the Galactic Plane ($z$$>$250~pc) may contain a greater fraction of large grains than the Milky Way average.

\end{abstract}

\section{Introduction}
\label{section_intro}
Dust is an integral component of the interstellar medium, infiltrating signals from both Galactic and extragalactic objects. Interstellar dust extincts and reddens background light, necessitating corrections in most astronomical observations. Dust grains range in size depending on the environment and are thought to follow a power law distribution as the result of formation and destruction processes \citep{Clayton_2003}. Precise corrections for dust depend on understanding not only its spatial distribution but also the grain size distribution and its extinction properties in all three dimensions. All-sky infrared dust maps, like those from \citet{Schlegel_1998} and \citet{Schlafly_2011}, provide a low resolution image of dust distribution in the Milky Way. \citet{Schlafly_2017} and \citet{Green_2019} have revealed dust populations as a function of Galactic position and distance. However, these dust maps are not capable of revealing how grain properties vary as a function of local environments nor have they completely constrained how those properties may change above the Plane.

Dust is often characterized by its extinction in magnitudes at V band ($A_{\rm{V}}$) and ratio of interstellar extinction to reddening ($R_{\rm{V}}$). The two quantities are related by $A_{\rm{V}}$$=$ $R_{\rm{V}}$$E(B-V)$, where $E(B-V)$ is the color excess or reddening. The average $R_{\rm{V}}$ in the Milky Way is $\approx$3.1 \citep{mw_rv}. Larger $R_{\rm V}$ indicates an abundance of large dust grains, while a smaller $R_{\rm V}$ indicates a shift toward smaller dust grains \citep{Mathis_1981}. The transition between dense and diffuse clouds at $A_{\rm{V}}$$\approx $3.2 mag. corresponds to a transition in $R_{\rm{V}}$ as well: dense clouds have large $R_{\rm{V}}$ (3.5--4.0), while diffuse clouds have small or near-average $R_{\rm{V}}$ \citep{whittet2001}. Large $R_{\rm{V}}$ clouds show a reduction in the number of small dust grains, possibly as a result of aggregation of small particles in dense environments \citep{Kim}. In diffuse cloud environments, ultraviolet photons may alter the grain size distribution in the opposite sense, leading to a small $R_{\rm{V}}$ through the destruction and fragmentation of large grains \citep{Aannestad_1995}. Many studies using coarse maps of interstellar reddening \citep[for example, ][]{Green_2019} to correct for dust assume $R_{\rm{V}}$$\approx$3.1 and therefore do not account for dust properties that may vary by sightline or even with distance along a single sightline. 

While dust properties like $A_{\rm{V}}$ and $R_{\rm{V}}$ can be measured with photometry and spectrophotometry, optical polarimetry of background stars provides another tool to probe the size, orientation, and alignment of grains. \citet{Hiltner} and \citet{Hall} first discovered that dust polarizes background starlight, which is a consequence of non-spherical dust grains aligned perpendicular to the interstellar magnetic field \citep{davis}. Radiative torques (RATs) have emerged as the dominant theory for grain alignment. RAT theory suggests that helical dust grains are spun up by an anisotropic radiation field so that their long axes are perpendicular to the magnetic field  \citep{Lazarian}. A larger fraction of dust grains are aligned and create a more polarized signal when the ambient radiation is along magnetic field lines \citep{Andersson2011}. Grain alignment is therefore likely enhanced near radiation sources having high specific intensity \citep{Andersson}. 

The mechanism of grain alignment depends on the size and composition of dust grains and the radiative environment. By studying the dark star-forming cloud $\rho$ Ophiuchi in far-infrared emission, \citet{Santos_2019} discovered that polarization fraction decreases with increasing column density (and hence, extinction), possibly due to increased magnetic field disorder or a loss of alignment efficiency deep within clouds. They proposed that warm dust grains at the outskirts of clouds are aligned through RAT while inner cold grains are shielded. While all grain alignment mechanisms fail at the center of dark clouds, those with smaller extinctions are expected to show more efficient alignment \citep{lazarian&goodman}. 

The high-space-velocity, hot, O9.2IV star $\zeta$~Ophiuchi makes an interesting laboratory for examining extinction and polarization along sightlines through a diffuse high-radiation environment. Because it is nearby, at moderate Galactic latitude (b=$+$24\degr), and has a well-characterized foreground \citep{serkowski, Tachihara_2000, liszt, choi, Siebenmorgen_2020}, the field of view around the star allows for relatively unconfused measurements of dust properties with distance. \textit{Hipparcos} parallax measurements place $\zeta$~Oph at a distance of 112~pc \citep{Hipparcos}, but more recent estimates from $Gaia$ suggest a distance of 182$^{+53}_{-34}$~pc \citep{Bailer_Jones_2018}. Because $\zeta$~Oph has an m$_{\rm G}$=2.46, it falls into a regime in which $Gaia$ distances may have large systematic errors \citep{Drimmel_2019}. However, \citet{Siebenmorgen_2020} used two other distance estimates, the \ion{Ca}{2} distance scale and the spectrophotometric distance, to calculate distances of 230$^{+8}_{-12}$~pc. We will adopt a distance of 182$^{+53}_{-34}$~pc for the remainder of this work. Figure \ref{rawpol} illustrates the $\zeta$~Oph region and reveals the prominent dust structures in infrared images from the \textit{Spitzer Space Telescope} (\textit{Spitzer}). Red/green/blue represent  24/8/4.5~$\mu$m respectively. $\zeta$~Oph is preceded by a high-surface-brightness bowshock nebula, first discovered by \citet{gull} in H$\alpha$ at a standoff distance of 299\arcsec\ or 0.26~pc at the adopted distance of 182~pc \citep{Kobulnicky_2019}. The high proper motion of $\zeta$~Oph is parallel to the symmetry axis of the bowshock nebula and is shown by the magenta vector in Figure~\ref{rawpol}. Fainter infrared dust structures also span the \textit{Spitzer} image. These striations align approximately perpendicular to the bowshock structure and parallel to the peculiar motion of $\zeta$~Oph. The relationship between the fainter striations in Figure \ref{rawpol} and the bowshock nebula is unclear. The striations are large and extend much further than the bowshock itself and well beyond the pictured field of view. These dust structures--- the striations and the bowshock--- are large in angular size, making them both interesting and easily accessible for investigating $A_{\rm{V}}$, $R_{\rm{V}}$, and polarization through this diffuse, high radiation environment.

This work investigates the reddening and polarization along lines of sight in the high-radiation, diffuse environment surrounding $\zeta$~Oph in order to identify and characterize dust populations in that field as a function of distance. Section \ref{section_observ} introduces new spectrophotometry from the Apache Point Observatory, optical polarimetry from the Wyoming Infrared Observatory, and data reduction. Section \ref{section_analysis} presents the spectral types, reddening estimates, and polarizations towards 27 background stars. Using available photometry in conjunction with our observations, we identify three primary dust structures located at various distances in the field of view. Section \ref{section_discuss} explores the three dust populations and their relationships to the projected magnetic field and infrared striations. In Section \ref{section_summary}, we summarize the conclusions from this investigation. 

\section{Observations and Data Reduction}
\label{section_observ}

We selected stars in close angular proximity to $\zeta$~Oph that were accessible to the observational limits of spectroscopy and polarimetry on a 2--4~m telescope. This resulted in twenty-seven target stars both directly behind the bowshock nebula and in the surrounding field, with magnitudes $m_f\leq 15.2$, where $m_f$ is the UCAC4 catalog optical photographic fit model magnitude \citep{Zacharias_2013}. Table~\ref{targets} lists basic data for the 27 targets. Column 1 provides the target number that is referenced throughout this paper. Column 2 lists the $Gaia$ identifier. Columns 3 and 4 provide the right ascension and declination of the target, column 5 gives the UCAC4 optical magnitude, and column 6 gives the V-band magnitude from the \textit{AAVSO Photometric All-Sky Survey} \citep[\textit{APASS};][]{Henden_2016yCat.2336....0H}. Where \textit{APASS} data were unavailable (targets \#14, \#15, and \#17), we use Panoramic Survey Telescope and Rapid Response System \citep[\textit{Pan-STARRS;}][]{chambers2019panstarrs1} photometry in g$^{\prime}$ and r$^{\prime}$ and the transformations from \citet{Jester2005} for the V-band magnitude. Column 7 gives the $Gaia$ distance from \citet{Bailer_Jones_2018}. Target distances appear bi-modal (as seen in subsequent Figures \ref{Av} and \ref{fig:fitz_dist}) and fall into two distinct groups: a ``near group'' (14 stars) with distances $\leq$600~pc and a ``far group'' (13 stars) with distances $\geq$1000~pc. 

\begin{table}[htb]
    \centering
    \caption{Targets observed in the $\zeta$~Ophiuchi field.}
    \label{targets}
    \begin{tabular}{@{}ccccccc@{}}
        \toprule
        Target & \textit{Gaia} ID & R.A. & Dec. & m$_{\rm{f}}$ & m$_{\rm{V}}$ & d \\
        & & (\degr)\ & (\degr)\ & (mag.) & (mag.) & (pc) \\
        \hline
        $\zeta$~Oph & 4337352305315545088 & 249.2875 & -10.5669 & ... & 2.56 & 182$_{-34}^{+53}$ \\
        1 & 4338102545907455744 & 249.0266 & -10.5774 & 13.18 & 13.08 & 436.1$_{-9.1}^{+7.9}$  \\
        2 & 4338102722003440000 & 249.0303 & -10.5656 & 13.01 & 13.21 & 2819$_{-287}^{+358}$  \\
        3 & 4338103271759259392 & 249.0668 & -10.5238 & 14.55 & 14.51 & 3019$_{-227}^{+265}$  \\
        4 & 4338102923864645120 & 249.1036 & -10.5442 & 11.84 & 11.90 & 224.0$_{-2.0}^{+1.1}$  \\
        5 & 4338108288281067264 & 249.1334 & -10.4531 & 13.85 & 14.05 & 423.3$_{-5.3}^{+4.7}$  \\
        6 & 4338103684076094464 & 249.1438 & -10.5379 & 13.33 & 13.38 & 252.2$_{-2.2}^{+0.8}$  \\
        7 & 4338108460079758336 & 249.1620 & -10.4341 & 14.68 & 14.94 & 8434$_{-1480}^{+2077}$  \\
        8 & 4338128972843559808 & 249.1805 & -10.4329 & 14.43 & 14.69 & 2420$_{-170}^{+197}$  \\
        9 & 4338104199472164352 & 249.2086 & -10.5013 & 13.86 & 13.94 & 435.5$_{-5.5}^{+4.6}$  \\
        10 & 4338103507980197248 & 249.2088 & -10.5421 & 12.74 & 13.13 & 1294$_{-63}^{+68}$  \\
        11 & 4338104405630603904 & 249.2136 & -10.4704 & 14.61 & 14.97 & 278.3$_{-2.5}^{+2.6}$  \\
        12 & 4338104405630602240 & 249.2170 & -10.4745 & 15.10 & 15.49 & 3247$_{-375}^{+482}$  \\
        13 & 4338104126455487488 & 249.2367 & -10.4871 & 12.23 & 12.32 & 427.6$_{-6.6}^{+6.4}$  \\
        14 & 4337353404830237056 & 249.2431 & -10.5256 & 14.67 & 14.72\tablenotemark{a} & 489.2$_{-7.1}^{+7.3}$  \\
        15 & 4337353370470494464 & 249.2637 & -10.5214 & 14.80 & 15.07\tablenotemark{a} & 2500$_{-190}^{+223}$  \\
        16 & 4337377181769178752 & 249.3271 & -10.4865 & 13.49 & 13.69 & 2350$_{-125}^{+138}$  \\
        17 & 4337376700732832512 & 249.3277 & -10.5117 & 14.31 & 14.56\tablenotemark{a} & 2537$_{-190}^{+222}$  \\
        18 & 4337372543204454144 & 249.3699 & -10.5763 & 13.26 & 13.33 & 307.5$_{-5.5}^{+4.5}$  \\
        19 & 4337376288415957120 & 249.3703 & -10.5152 & 14.91 & 15.21 & 4645$_{-628}^{+842}$  \\
        20 & 4337376528934134144 & 249.3904 & -10.4833 & 14.79 & 14.90 & 2616$_{-190}^{+222}$  \\
        21 & 4337376460214656896 & 249.3975 & -10.4786 & 14.50 & 14.75 & 2299$_{-204}^{+246}$ \\
        22 & 4337372985583902848 & 249.4100 & -10.5507 & 15.17 & 15.44 & 4790$_{-709}^{+975}$  \\
        23 & 4337347598034360704 & 249.4194 & -10.6498 & 13.39 & 13.51 & 552.4$_{-6.4}^{+5.6}$  \\
        24 & 4337372852442084096 & 249.4617 & -10.5519 & 12.16 & 12.20 & 202.7$_{-1.7}^{+1.3}$  \\
        25 & 4337359452144088448 & 249.4686 & -10.6339 & 13.39 & 13.42 & 490.4$_{-5.4}^{+4.6}$  \\
        26 & 4337371787290173952 & 249.5049 & -10.5797 & 13.04 & 13.06 & 581.7$_{-13.7}^{+13.3}$  \\
        27 & 4337375218965869568 & 249.5108 & -10.4688 & 12.90 & 12.95 & 526.5$_{-10.5}^{+10.5}$  \\
        \hline
    \end{tabular}
    \tablenotetext{a}{\textit{APASS} photometry was not available. \textit{Pan-STARRS} photometry is used with transformations from \citet{Jester2005}.}
\end{table}

\subsection{APO Spectrophotometry}
We obtained spectra of the 27 targets using the Apache Point Observatory (APO) Dual Imaging Spectrograph (DIS) on 2019 May 2 and 3 (21 targets) and 2021 March 30 (six targets). The two channels of the spectrograph covered 4000--5240~\AA\ with a reciprocal dispersion of 0.62~\AA\ pix$^{-1}$ and 5720--6880~\AA\ with a reciprocal dispersion of 0.58~\AA\ pix$^{-1}$, both using a 1200 lines~mm$^{-1}$ grating. Observations were taken with 1\farcs{5}--2\arcsec\ slits oriented at the parallactic angle. The seeing average was near 2\arcsec\ over 1.3--1.7 airmasses. Exposure times ranged from 300~s to 600~s with 1--2 exposures per star. The continuum SNR per pixel ranged from 16--37 at 4500~\AA\ and 40--71 at 6200~\AA. The data were wavelength calibrated using HeNeAr exposures every 1--2 hours. Observations of spectrophotometric standard stars HZ~44, BD+33~2642, and BD+28~4211 from \citet{Oke_1990} over airmasses 1.07--2.06 were used for flux calibration.

The spectra were processed in the standard manner with IRAF \citep{Tody_1986SPIE..627..733T} tasks to remove bias level using overscan regions on each exposure and correct for pixel-to-pixel variations with continuum flat fields. We corrected for atmospheric extinction over a range of airmasses with standard star data. The extinction-corrected standard stars differ from the \citet{Oke_1990} spectra by no more than 7\% at extreme blue end and smaller percentages at the extreme red end. 

\subsection{WIRO Polarimetry}
Polarimetric observations of the 27 stars were made 2019 April 26--28, May 4, and June 6 using the OptiPol optical polarimeter \citep{optipol} at the Cassegrain focus of the Wyoming Infrared Observatory (WIRO) 2.3~m telescope.  The instrument employs a Wollaston prism to image orthogonal polarizations simultaneously onto a 1024$^2$ Apogee Alta F16M with Kodak KAF-16803 CCD array with 9~$\mu$m pixels at a scale of 0\farcs{12} pix$^{-1}$ (when on-chip 4$\times$4 binning is used), yielding a 40\arcsec$\times$80\arcsec\  field of view.  A sequence of exposures was obtained in the V-band filter for each target at four angular positions (0\degr, 22.5\degr, 45\degr, and 67.5\degr) of a half-wave plate. Exposure times ranged from 6--200 seconds, according to target brightness. Unpolarized standards BD$+$33 2642, BD$+$28 4211 \citep{pol_standards}, and  HD~154892 \citep{pol_standards2} were observed to determine the instrumental polarization for each night. 

The data from each position of the waveplate were processed in a standard manner using IRAF \citep{Tody_1986SPIE..627..733T} to remove contributions from dark current, bias level, and pixel-to-pixel sensitivity variations. We used the {\tt LACosmics} python script \citep[based on algorithms from and developed by][]{Vandokkum} to clean cosmic rays from the data. We used a custom python code to perform aperture photometry on each image and compute the normalized \textit{q} and \textit{u} Stokes parameters using the method from \citet{Tinbergen}. 

Polarized standard stars HD~155197, HD~161056, and VI Cyg 12 \citep{pol_standards} were used to determine an instrumental position angle offset of 81.5$^{\circ}$ and verify the percent polarization. Average instrumental polarization was small ($q$=-0.20, $u$=0.19). The instrumental polarization was subtracted in the $q$--$u$ plane, and the calibrated Stokes parameters were calculated. Using these quantities, the percent polarization ($P_{\rm{V}}'$) and the position angle East of North (P.A.) are given by the standard relations: $P_{\rm{V}}'$=$\sqrt{q^2 + u^2}$ and P.A.$=$$\frac{1}{2}\tan^{-1}\frac{u}{q}$. To correct the data for positive bias, we used the relationship that $P_{\rm{V}}$=$\sqrt{P_{\rm{V}}'^{2}-\sigma_{P}^{2}}$, where $\sigma_{P}$ is the polarization uncertainty propagated from errors in $q$ and $u$  \citep{Wardle_1974ApJ...194..249W}. 

Table~\ref{poltab} reports the polarimetric results. Column 1 lists the target number corresponding to Table~\ref{targets}, columns 2 and 3 list the $q$ and $u$ values, column 4 lists the percent polarizations, $P_{\rm{V}}$,  and 1$\sigma$ uncertainties, and column 5 lists the position angles, P.A., and 1$\sigma$ uncertainties. 

\begin{table}[H]
    \centering
    \caption{Polarization data for the target stars in Table~\ref{targets}.}
    \label{poltab}
    \begin{tabular}{@{}ccccccccc@{}}
        \toprule
        Target & \textit{q} (\%) & \textit{u} (\%) & $P_{\rm{V}}$ (\%)  & P.A. (\degr) & \textit{q$_f$} (\%) & \textit{u$_f$} (\%) & $P_{\rm{V f}}$ (\%)  & P.A.$_f$ (\degr) \\
        \hline
        $\zeta$~Oph & $-0.45 \pm 0.04$ & $-1.37 \pm 0.04$ & $1.44 \pm 0.04$ & $126.0 \pm 0.8$ & ... & ... & ... & ... \\
        1 & $ -1.04 \pm 0.28 $ & $ -0.36 \pm 0.27 $ & $ 1.06 \pm 0.28 $ & $ 99.5 \pm 7.1 $ & $ -0.60 \pm 0.28 $ & $ 1.01 \pm 0.27 $ & $ 1.14 \pm 0.27 $ & $ 60.3 \pm 6.8 $ \\ 
        2 & $ -1.26 \pm 0.23 $ & $ -0.61 \pm 0.25 $ & $ 1.38 \pm 0.23 $ & $ 102.9 \pm 5.0 $ & $ -0.82 \pm 0.23 $ & $ 0.76 \pm 0.25 $ & $ 1.09 \pm 0.24 $ & $ 68.5 \pm 6.2 $ \\ 
        3 & $ -1.78 \pm 0.21 $ & $ -1.00 \pm 0.21 $ & $ 2.03 \pm 0.21 $ & $ 104.7 \pm 3.0 $ & $ -1.34 \pm 0.21 $ & $ 0.37 \pm 0.21 $ & $ 1.37 \pm 0.21 $ & $ 82.3 \pm 4.3 $ \\ 
        4 & $ -0.60 \pm 0.10 $ & $ -1.80 \pm 0.10 $ & $ 1.89 \pm 0.10 $ & $ 125.8 \pm 1.5 $ & $ -0.15 \pm 0.10 $ & $ -0.43 \pm 0.10 $ & $ 0.45 \pm 0.10 $ & $ 125.1 \pm 6.3 $ \\ 
        5 & $ -1.41 \pm 0.18 $ & $ -1.27 \pm 0.18 $ & $ 1.89 \pm 0.18 $ & $ 111.0 \pm 2.7 $ & $ -0.96 \pm 0.18 $ & $ 0.10 \pm 0.18 $ & $ 0.95 \pm 0.18 $ & $ 87.0 \pm 5.3 $ \\ 
        6 & $ -0.81 \pm 0.42 $ & $ -1.82 \pm 0.25 $ & $ 1.97 \pm 0.29 $ & $ 123.0 \pm 5.7 $ & $ -0.37 \pm 0.42 $ & $ -0.45 \pm 0.25 $ & $ 0.48 \pm 0.33 $ & $ 115.5 \pm 17.9 $ \\
        7 & $ -1.68 \pm 0.25 $ & $ -0.97 \pm 0.24 $ & $ 1.92 \pm 0.25 $ & $ 105.0 \pm 3.6 $ & $ -1.23 \pm 0.25 $ & $ 0.40 \pm 0.24 $ & $ 1.27 \pm 0.25 $ & $ 81.0 \pm 5.3 $ \\ 
        8 & $ -1.38 \pm 0.29 $ & $ -1.75 \pm 0.28 $ & $ 2.21 \pm 0.28 $ & $ 115.9 \pm 3.7 $ & $ -0.93 \pm 0.29 $ & $ -0.38 \pm 0.28 $ & $ 0.97 \pm 0.29 $ & $ 101.1 \pm 8.0 $ \\ 
        9 & $ -1.54 \pm 0.16 $ & $ -0.93 \pm 0.16 $ & $ 1.79 \pm 0.16 $ & $ 105.6 \pm 2.6 $ & $ -1.10 \pm 0.16 $ & $ 0.44 \pm 0.16 $ & $ 1.17 \pm 0.16 $ & $ 79.1 \pm 3.9 $ \\ 
        10 & $ -1.93 \pm 0.23 $ & $ -1.08 \pm 0.22 $ & $ 2.20 \pm 0.23 $ & $ 104.6 \pm 2.9 $ & $ -1.48 \pm 0.23 $ & $ 0.29 \pm 0.22 $ & $ 1.50 \pm 0.23 $ & $ 84.5 \pm 4.2 $ \\ 
        11 & $ -2.07 \pm 0.35 $ & $ -0.99 \pm 0.35 $ & $ 2.27 \pm 0.35 $ & $ 102.8 \pm 4.4 $ & $ -1.62 \pm 0.35 $ & $ 0.38 \pm 0.35 $ & $ 1.63 \pm 0.35 $ & $ 83.4 \pm 6.0 $ \\ 
        12 & $ -1.05 \pm 0.25 $ & $ -1.89 \pm 0.24 $ & $ 2.15 \pm 0.24 $ & $ 120.5 \pm 3.3 $ & $ -0.60 \pm 0.25 $ & $ -0.52 \pm 0.24 $ & $ 0.76 \pm 0.25 $ & $ 110.3 \pm 8.8 $ \\ 
        13 & $ -1.03 \pm 0.07 $ & $ -0.85 \pm 0.07 $ & $ 1.33 \pm 0.07 $ & $ 109.8 \pm 1.5 $ & $ -0.58 \pm 0.07 $ & $ 0.52 \pm 0.07 $ & $ 0.78 \pm 0.07 $ & $ 69.2 \pm 2.6 $ \\ 
        14 & $ -2.13 \pm 0.42 $ & $ -0.75 \pm 0.42 $ & $ 2.22 \pm 0.42 $ & $ 99.7 \pm 5.3 $ & $ -1.68 \pm 0.42 $ & $ 0.62 \pm 0.42 $ & $ 1.75 \pm 0.42 $ & $ 79.9 \pm 6.7 $ \\ 
        15 & $ -1.55 \pm 0.29 $ & $ 0.02 \pm 0.31 $ & $ 1.52 \pm 0.29 $ & $ 89.6 \pm 5.7 $ & $ -1.10 \pm 0.29 $ & $ 1.39 \pm 0.31 $ & $ 1.75 \pm 0.30 $ & $ 64.2 \pm 4.8 $ \\ 
        16 & $ -1.75 \pm 0.25 $ & $ -0.89 \pm 0.22 $ & $ 1.95 \pm 0.24 $ & $ 103.5 \pm 3.3 $ & $ -1.30 \pm 0.25 $ & $ 0.48 \pm 0.22 $ & $ 1.37 \pm 0.25 $ & $ 79.9 \pm 4.6 $ \\ 
        17 & $ -1.68 \pm 0.24 $ & $ -0.34 \pm 0.23 $ & $ 1.70 \pm 0.24 $ & $ 95.7 \pm 3.9 $ & $ -1.23 \pm 0.24 $ & $ 1.03 \pm 0.23 $ & $ 1.59 \pm 0.24 $ & $ 70.1 \pm 4.2 $ \\ 
        18 & $ -1.39 \pm 0.39 $ & $ 0.48 \pm 0.39 $ & $ 1.42 \pm 0.39 $ & $ 80.5 \pm 7.6 $ & $ -0.94 \pm 0.39 $ & $ 1.85 \pm 0.39 $ & $ 2.04 \pm 0.39 $ & $ 58.5 \pm 5.4 $ \\ 
        19 & $ -1.63 \pm 0.40 $ & $ -0.72 \pm 0.36 $ & $ 1.74 \pm 0.39 $ & $ 101.9 \pm 5.9 $ & $ -1.18 \pm 0.40 $ & $ 0.65 \pm 0.36 $ & $ 1.29 \pm 0.39 $ & $ 75.6 \pm 7.8 $ \\ 
        20 & $ -1.20 \pm 0.29 $ & $ -0.82 \pm 0.28 $ & $ 1.42 \pm 0.29 $ & $ 107.2 \pm 5.6 $ & $ -0.75 \pm 0.29 $ & $ 0.55 \pm 0.28 $ & $ 0.89 \pm 0.29 $ & $ 72.0 \pm 8.7 $ \\ 
        21 & $ -1.33 \pm 0.30 $ & $ -0.23 \pm 0.29 $ & $ 1.32 \pm 0.30 $ & $ 94.9 \pm 6.2 $ & $ -0.88 \pm 0.30 $ & $ 1.14 \pm 0.29 $ & $ 1.41 \pm 0.29 $ & $ 63.9 \pm 5.9 $ \\ 
        22 & $ -2.04 \pm 0.40 $ & $ 0.66 \pm 0.42 $ & $ 2.11 \pm 0.40 $ & $ 81.0 \pm 5.6 $ & $ -1.60 \pm 0.40 $ & $ 2.03 \pm 0.42 $ & $ 2.55 \pm 0.41 $ & $ 64.1 \pm 4.5 $ \\ 
        23 & $ -1.22 \pm 0.15 $ & $ -0.30 \pm 0.15 $ & $ 1.25 \pm 0.15 $ & $ 96.9 \pm 3.4 $ & $ -0.77 \pm 0.15 $ & $ 1.07 \pm 0.15 $ & $ 1.31 \pm 0.15 $ & $ 63.0 \pm 3.3 $ \\ 
        24 & $ -0.78 \pm 0.07 $ & $ -0.75 \pm 0.07 $ & $ 1.08 \pm 0.07 $ & $ 111.9 \pm 1.9 $ & $ -0.34 \pm 0.07 $ & $ 0.62 \pm 0.07 $ & $ 0.70 \pm 0.07 $ & $ 59.2 \pm 2.9 $ \\ 
        25 & $ -0.58 \pm 0.53 $ & $ -0.01 \pm 0.46 $ & $ 0.24 \pm 0.53 $ & $ 90.5 \pm 22.7 $ & $ -0.13 \pm 0.53 $ & $ 1.36 \pm 0.46 $ & $ 1.29 \pm 0.46 $ & $ 47.8 \pm 11.1 $ \\ 
        26 & $ -1.89 \pm 0.23 $ & $ -0.15 \pm 0.16 $ & $ 1.88 \pm 0.23 $ & $ 92.3 \pm 2.4 $ & $ -1.44 \pm 0.23 $ & $ 1.22 \pm 0.16 $ & $ 1.88 \pm 0.20 $ & $ 69.9 \pm 2.9 $ \\ 
        27 & $ -1.45 \pm 0.12 $ & $ -0.42 \pm 0.12 $ & $ 1.50 \pm 0.12 $ & $ 98.1 \pm 2.3 $ & $ -1.00 \pm 0.12 $ & $ 0.95 \pm 0.12 $ & $ 1.38 \pm 0.12 $ & $ 68.3 \pm 2.5 $ \\ 
        \hline
    \end{tabular}
    \tablenotetext{ }{$_f$ represents polarization values after subtracting $\zeta$~Oph in the q--u plane.}
\end{table}

\begin{figure}
    \centering
    \includegraphics[width=7in]{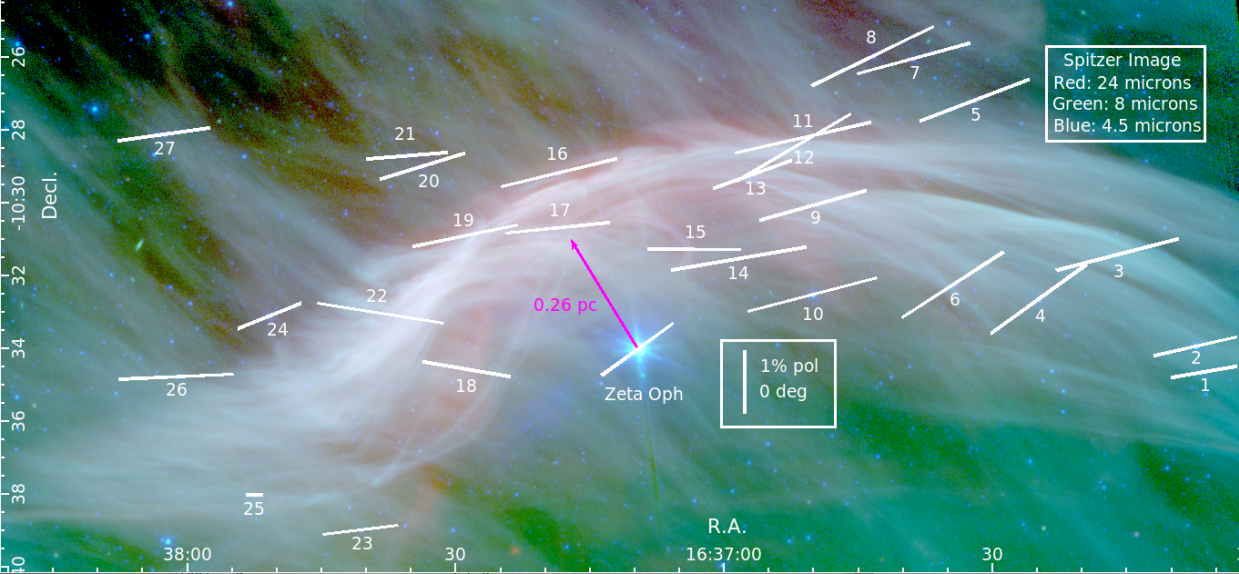}
    \caption{The $\zeta$~Oph field from \textit{Spitzer} 24~$\mu$m, 8~$\mu$m, and 4.5~$\mu$m images. The bowshock nebula precedes the star and has a high surface brightness. At a distance of 182~pc, the apex of the nebula is a projected 0.26 pc from $\zeta$~Oph. The magenta vector shows this distance and the direction of the star's motion. Fainter dust striations span the entire field of view and are best traced in green (8~$\mu$m). White vectors show $P_{\rm{V}}$ and P.A. relative to the $P_{\rm{V}}$=1\% and P.A.=0\degr\ legend. Numbers next to each vector provide the target identifier as in Table~\ref{targets}.}
    \label{rawpol}
    \includegraphics[width=7in]{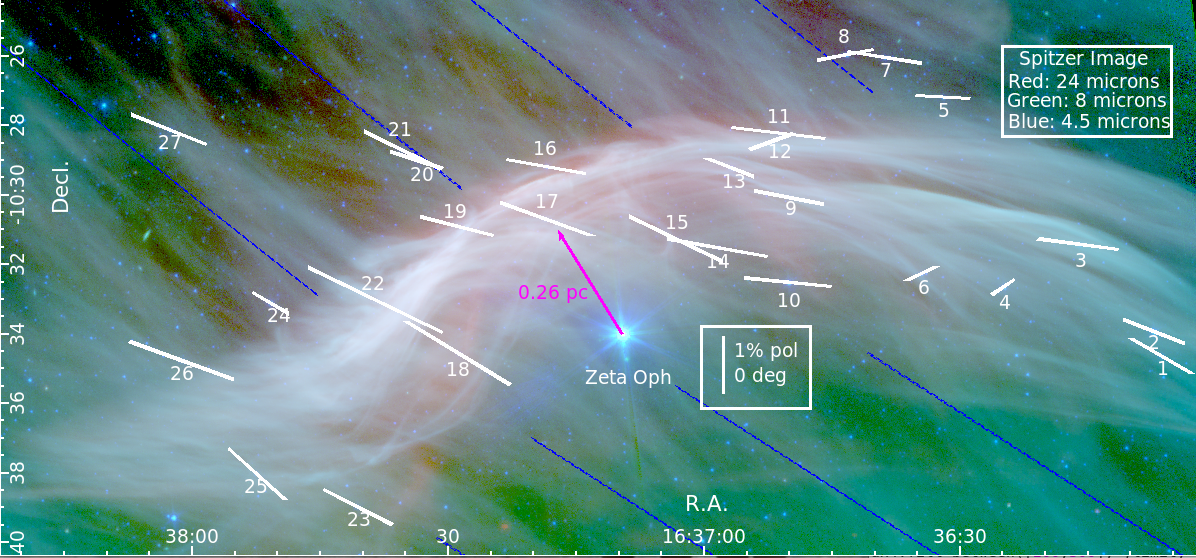}
    \caption{Foreground-subtracted polarization vectors, $P_{\rm{V f}}$ and PA$_{\rm{f}}$, overlaid on the same infrared images as in Figure \ref{rawpol}. Blue dashed lines show the orientation of the 8~$\mu$m dust striations above and below the bowshock as traced by the Rolling Hough Transform from \citet{Clark_2014ApJ...789...82C}, detailed in Section \ref{section_striations}.}
    \label{subpol}
\end{figure}

Figure~\ref{rawpol} shows the V-band polarization vectors overlaid on a 32\arcmin$\times$18\arcmin\ version of the \textit{Spitzer} field surrounding $\zeta$~Oph. The orientation of the numbered white vectors demonstrates the polarization position angle for each target in Table~\ref{targets}. The length of the vectors shows the percent polarization. The legend illustrates 1\% polarization at PA=0\degr. We also show the polarization of $\zeta$~Oph at V band to be 1.44\% at PA=126$\degr$ \citep{serkowski}. 

A majority of the stars in Figure~\ref{rawpol} show similar position angles and polarization percentages. The mean P.A. for the 27 targets is 103\degr\ with a standard deviation of 11\degr\ and the mean $P_{\rm{V}}$ is 1.68\% with a standard deviation of 0.46\%. The mean position angle of the polarization vectors is smaller than that of $\zeta$~Oph, but the mean percent polarization is larger than that of $\zeta$~Oph \citep{serkowski}. Targets at similar distances to $\zeta$~Oph, like \#4 (224~pc) and \#6 (252~pc), are in better agreement with the position angle of $\zeta$~Oph than targets at larger distances and have larger-than-average polarizations. Other stars at modest distances (278--500~pc; \#11, \#13, \#14) and beyond exhibit a rotation in position angle relative to the nearest stars and have a notable uniformity in $P_{\rm{V}}$ and P.A. 

To examine the polarization beyond $\zeta$~Oph, we subtracted the polarization of $\zeta$~Oph itself from each of our polarization measurements in the \textit{q}--\textit{u} plane. For $\zeta$~Oph, \textit{q}=$-$0.445 and \textit{u}=$-$1.37 from \citet{serkowski}. When added in quadrature, the errors from our measurements dominate over the errors in the polarization of $\zeta$~Oph. Table \ref{poltab} columns 6, 7, 8, and 9 provides the foreground-subtracted quantities, $q_{\rm f}$, $u_{\rm f}$, $P_{\rm{V f}}$, and P.A.$_{\rm{f}}$.\footnote{$q_{\rm f}$=$q$-(-0.445) and $u_{\rm f}$=$u$-(-1.37).  $P_{\rm{V f}}$ and P.A.$_{\rm{f}}$ are constructed in the usual manner from these quantities. } Using the same notation as Figure \ref{rawpol}, Figure~\ref{subpol} presents the foreground-subtracted polarization vectors on the infrared \textit{Spitzer} image. This subtraction results in much smaller $P_{\rm{V f}}$ for the three nearest stars (targets \#4, \#6, and \#24) and an abrupt change in P.A. for the rest of the field. The foreground-subtracted mean values are $P_{\rm{V f}}$=1.29\% with a dispersion of 0.47\% and P.A.$_{\rm{f}}$=77.2\degr\ with a dispersion of 17.8\degr. 

\section{Analysis}
\label{section_analysis}

\subsection{Extinction Estimates}

In order to investigate the interstellar extinction throughout the field of view, we estimated $A_{\rm{V}}$ using four different methods. Three of these methods, $A_{\rm{V}}$(spectra), $A_{\rm{V}}$(phot) and $A_{\rm{V}}$(RJCE), assume $R_{\rm{V}}$=3.1 and are described in detail and compared below. The fourth, complementary but separate method allows $R_{\rm{V}}$ to deviate from the Milky Way average. The combination of these extinction estimates allows us to examine the dust properties as a function of distance. 

\subsubsection{Determining Target Spectral Types}

As a prerequisite to calculating extinction, we first determined the spectral types of the target stars. We compared our continuum-normalized spectra to continuum-normalized PHOENIX stellar models \citep{Husser_2013} to determine approximate T$_{\rm eff}$ by visual examination. As another check on the spectral types, we used the spectral classification code {\tt PyHammer} \citep{pyhammer}. Visual inspection agreed with the {\tt PyHammer} results within about 300~K. 

\begin{figure}
    \centering
    \includegraphics[width=5in]{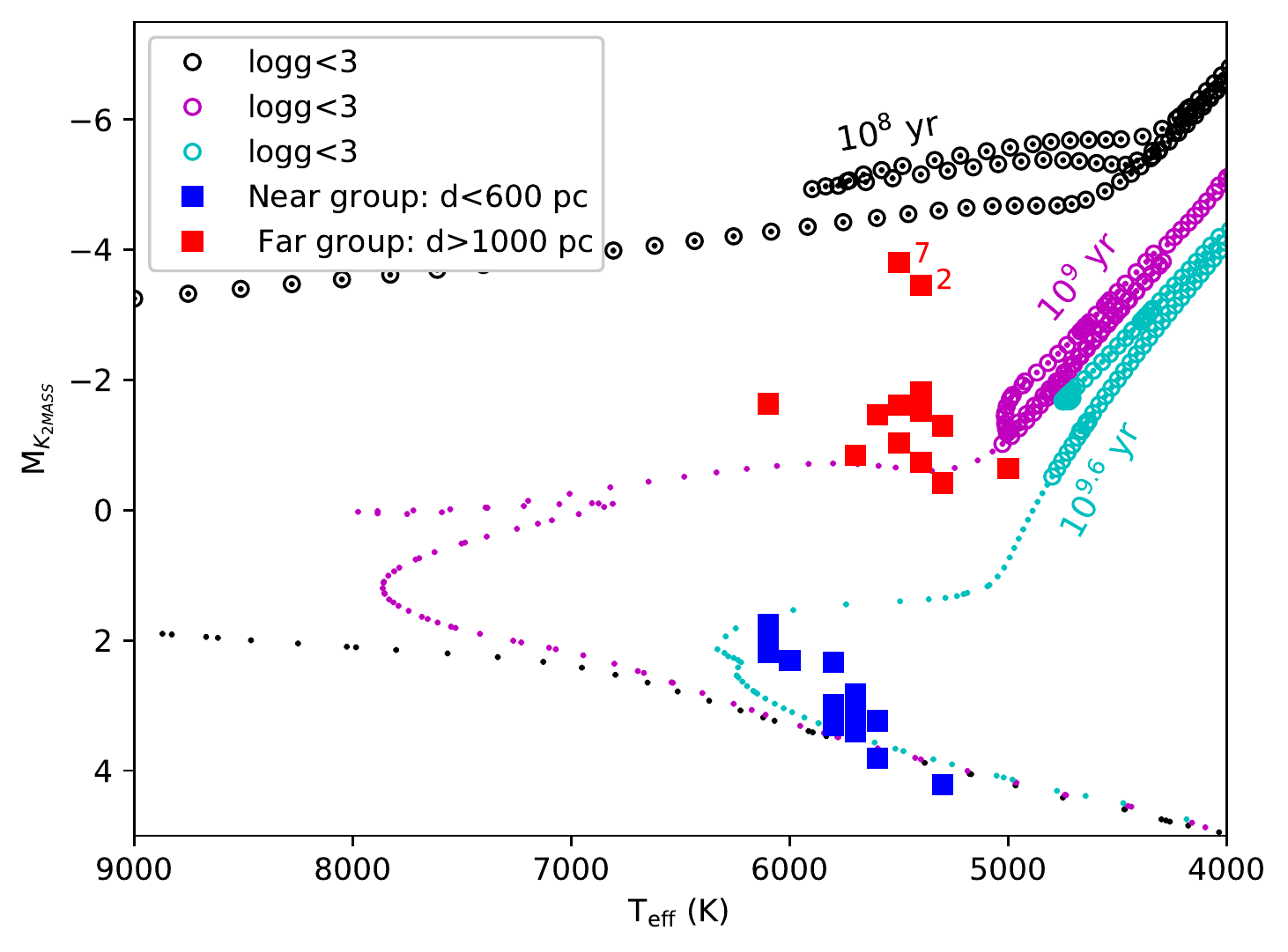}
    \caption{Absolute K magnitude versus effective temperature for the near group of background stars (blue squares) and the far group (red squares).  Colored tracks show solar metallicity PARSEC \citep{Bressnan2012} isochrones at ages of 10$^{9.6}$/10$^9$/10$^8$ yr, as labeled. Large symbols denote the approximate onset of giants at $\log$g$<$3. The near group is consistent with near-main-sequence F--G stars while the far group consists of post-main-sequence giants.}
    \label{fig:HRD}
\end{figure}

While our optical spectra provide an estimate of the effective temperature for the target stars, their positions on the H-R diagram offer additional insight regarding their evolutionary status (i.e. $\log$~$g$).  Figure~\ref{fig:HRD} plots the absolute K magnitude (as computed from the distance modulus) versus T$_{\rm eff}$ for the near group (blue squares) and the far group (red squares).  Cyan/magenta/black tracks show 10$^{9.6}$/10$^9$/10$^8$~yr isochrones from \citet[][PARSEC, version 1.2S, adopting solar metallicity]{Bressnan2012}\footnote{As computed using the web interface http://stev.oapd.inaf.it/cgi-bin/cmd on 2020 December 15.}. The main sequence emanates from the lower right corner of the plot. The main-sequence turnoffs are visible. The tracks change from points to encircled points as the stars evolve into giants ($\log$~$g$ falls below 3). The near group of stars at distances $d$$<$600 pc and $z$ heights $\lesssim$240 pc above the Plane are consistent with near-main-sequence F--G V--IV stars (the track of small cyan points), for which we may reasonably adopt $\log$~$g$=4.0 and metallicity [M/H]=0, appropriate for the Galactic disk population. The far group of targets at distances $d$$>$1000 pc and $z$ heights $>$400 pc are consistent with post-main-sequence giants with ages $<$1 Gyr (the track of small magenta points as they transition into encircled points).  At such young ages, the far group cannot---despite their larger $z$ distances---be metal-poor thick disk or halo stars.  We adopt $\log$~$g$=3.0 and metallicity [M/H]=0 for these objects.  For the two most luminous objects, 2 and 7, which are also among the most distant in our sample, we adopt $\log$~$g$=2.0. We created similar plots with reduced-metallicity ($-$0.5 dex) isochrones for comparison and found metallicity has negligible impact on the inferred ages and gravities.

Given these fundamental parameters ($T_{\rm eff}$, $\log$~$g$, [M/H]), we assigned a PHOENIX model spectrum to represent the intrinsic spectral energy distribution needed to characterize the interstellar extinction along each sightline. Table~\ref{exttab} lists target number in column 1, adopted effective temperature in column 2, adopted $\log$~$g$ in column 3, and corresponding approximate spectral type in column 4. Our targets are F, G, and K stars that correspond to PHOENIX model temperatures in the range 5000--6100~K, based on the effective temperature versus spectral type sequence from \citet{Pecaut_2013}. By visual comparison of PHOENIX models to spectra, we estimate that our classifications are accurate to $\pm$300~K, or approximately 3--4 subtypes.

\subsubsection{Fitting for $A_{\rm{V}}$(spectra) with APO Spectra}

As an initial extinction estimate, we calculated $A_{\rm{V}}$ using the APO spectra. We smoothed the stellar model templates and the data to ten broad wavelength intervals, as only the overall shape of the spectrum is needed to fit for extinction. We normalized the stellar model from Table~\ref{exttab} and the flux-calibrated spectra at the longest wavelength interval, but the derived extinctions remain the same with any normalization. We then divided the spectra by the corresponding PHOENIX model to obtain a curve that reflects the wavelength-dependent interstellar extinction, $A_{\rm{\lambda}}$. The curve was then fit using the \citet[][equations 1, 2, and 3]{CCM89} optical reddening parameterization with a fixed  $R_{\rm{V}}$=3.1 to obtain $A_{\rm{V}}$(spectra). The resulting range of $A_{\rm{V}}$(spectra) is 0.79--3.27 magnitudes, and values are tabulated in column 5 of Table~\ref{exttab}. Uncertainties on $A_{\rm{V}}$(spectra) (also shown in Table~\ref{exttab} column 5) are dominated by uncertainty on the spectral type. We performed the same fitting process using stellar models 300~K below and above the chosen model and used the dispersion in the resultant values of $A_{\rm{V}}$(spectra) as the 1$\sigma$ uncertainties. The average uncertainty in $A_{\rm{V}}$(spectra) is $\approx$0.3 mag. 

\subsubsection{Fitting for $A_{\rm{V}}$(phot) with Publicly Available Photometry}

Because our spectra cover only the optical range of the spectrum, we also estimated the extinction with photometry in the optical and IR. We performed the same extinction calculation with \textit{APASS} photometric data in B, V, g$^{\prime}$r$^{\prime}$i$^{\prime}$z$^{\prime}$y$^{\prime}$, \textit{Two Micron All Sky Survey} (\textit{2MASS}) JHK, and \textit{Spitzer} 3.6~$\mu$m and 4.5~$\mu$m magnitudes as catalogued in the NASA/IPAC Infrared Science Archive (\textit{IRSA}) database. \textit{APASS} data are available for 24 of the 27 target stars (missing are \#14, \#15, and \#17 on account of their proximity to $\zeta$~Oph). For those targets, we used photometric data from \textit{Pan-STARRS}. 

We normalized the stellar model from Table~\ref{exttab} to the photometry at 4.5 $\mu$m, then performed synthetic photometry on each model as described by \citet{Girardi_2002} using published filter transmission curves specific to each bandpass. We divided the fluxes of the photometric measurements in each band by the synthetic stellar model fluxes, then fitted for $A_{\rm{V}}$(phot) using the same approach as above, including the \citet{CCM89} prescriptions for the optical and NIR. Column 6 of Table~\ref{exttab} lists $A_{\rm{V}}$(phot) derived from the photometric data. The resulting range of $A_{\rm{V}}$(phot) is 0.69--2.61 magnitudes. Uncertainties again are dominated by uncertainties on the spectral type and are calculated in the same manner. The average error in $A_{\rm{V}}$(phot) is $\approx$0.2 mag, somewhat smaller than that of the spectra on account of the larger wavelength range used.   

\begin{figure}
    \centering
    \includegraphics[width=5in]{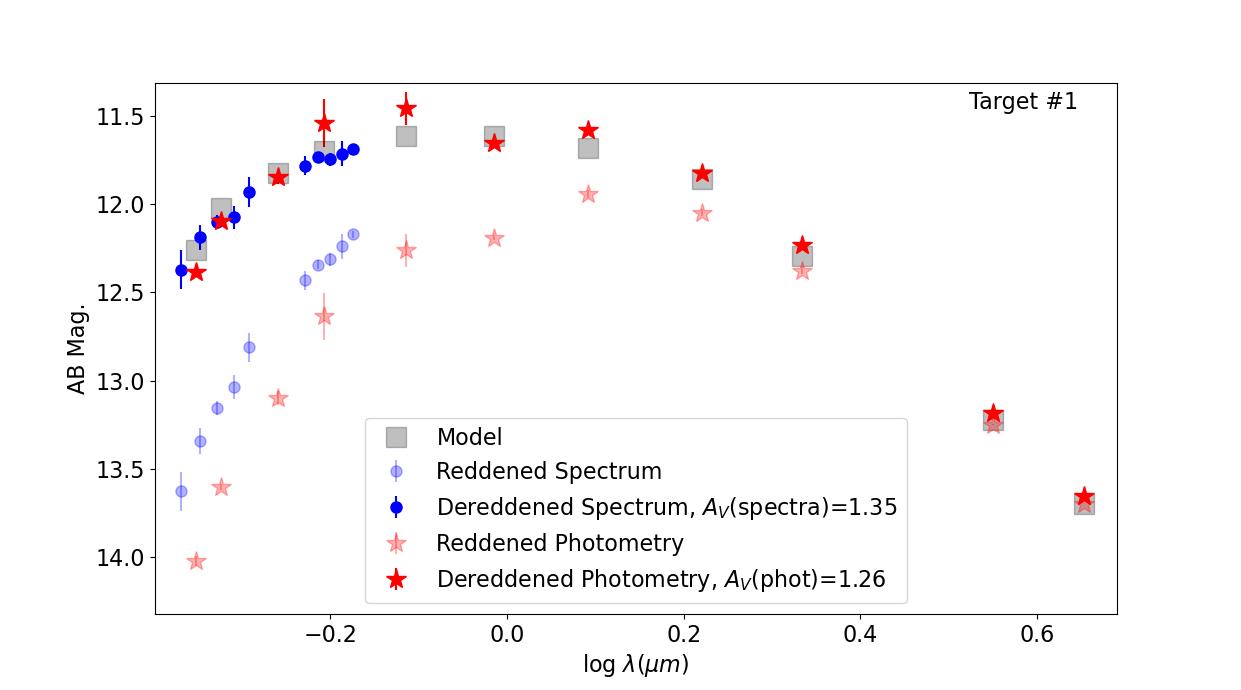}
    \caption{Reddening correction and extinction estimation for target \#1. The gray squares show the smoothed stellar model (6000 K, $\log$~$g$=4.0) corresponding to the F9 spectral type. Blue circles show the reddened APO spectrum (translucent) and APO spectrum after dereddening by $A_{\rm{V}}$(spectra)$=$1.35$^{+0.25}_{-0.26}$ and $R_{\rm{V}}$=3.1 (bold). Red stars show the reddened photometry (translucent) and photometry after dereddening by $A_{\rm{V}}$(phot)=$1.26^{+0.18}_{-0.20}$ and $R_{\rm{V}}$=3.1 (bold).}
    \label{extinction_correction}
\end{figure}

Figure~\ref{extinction_correction} shows an example of the technique used to estimate interstellar extinctions for target \#1. The gray squares show the stellar model at each bandpass (F9, 6000 K, $\log$~$g$=4.0) in AB magnitudes. The translucent blue circles show the reddened APO spectrum smoothed to ten overall bins. Errors represent the standard deviation in each bin. The translucent red stars show the reddened photometry, which has errors directly from \textit{IRSA}, \textit{APASS}, and \textit{Pan-STARRS} that are often too small to be visible in the figure. The spectrum and photometry differ slightly in AB mag prior to dereddening. This can be explained with an arbitrary gray shift resulting from clouds affecting the spectrophotometry during non-photometric conditions. However, the shape of both the reddened spectra and photometry is always very similar, and that shape determines the reddening and extinction. The bold blue circles show the spectrum after dereddening by $A_{\rm{V}}$(spectra)$=$1.35$^{+0.25}_{-0.26}$ and $R_{\rm{V}}$=3.1, and the bold red stars show the photometry after dereddening by $A_{\rm{V}}$(phot)=$1.26^{+0.18}_{-0.20}$ and $R_{\rm{V}}$=3.1. The two estimates agree within their uncertainties and provide a good match to the model template. 

\subsubsection{Calculating $A_{\rm{V}}$(RJCE) with the Rayleigh-Jeans Color Excess Method}

The Rayleigh-Jeans color excess (RJCE) method \citep{Majewski_2011} also provides an estimation for $A_{\rm{V}}$ based on H~band and 4.5~$\mu$m photometry from \textit{2MASS} and \textit{Spitzer} \citep{2mass, IRAC}. The H$-$4.5~$\mu$m color is a robust measure of extinction regardless of temperature, metallicity, or surface gravity for the ubiquitous F--M stars that populate most Galactic fields. Using the relationships $A_{\rm {K}}$=0.918(H$-$[4.5$\mu$]$-$0.08) and $A_{\rm{V}}$=8.82$A_{\rm{K}}$ for $R_{\rm{V}}$=3.1, we determined $A_{\rm{V}}$(RJCE) and uncertainties, as listed in column 7 of Table~\ref{exttab}. $A_{\rm{V}}$(RJCE) ranges from 0.55--2.54~mag with typical uncertainties $\approx$0.2 mag.

\begin{table}[ht]
    \centering
    \caption{Spectral types and extinction values for the target stars in Table~\ref{targets}.}
    \label{exttab}
    \begin{tabular}{@{}cccccccc@{}}
        \toprule
        Target & T$_{\rm eff}$ & $\log$~$g$  & Sp. Type & \multicolumn{3}{c}{$A_{\rm{V}}$ (mag.) assuming $R_{\rm{V}}$ = 3.1}  & $E(B-V)$\\
        &(K)&  & & $A_{\rm{V}}$(spectra) & $A_{\rm{V}}$(phot) & $A_{\rm{V}}$(RJCE) & (mag.)\\
        \hline
        1 & 6000 & 4.00 & F9 & 1.35 $_{- 0.26 }^{+ 0.25 }$ &   1.26 $_{- 0.20 }^{+ 0.18 }$ &   1.43 $\pm$ 0.20  &  0.49 $_{- 0.08 }^{+ 0.08 }$ \\ 
        2 & 5400 & 2.50 & G9 & 3.27 $_{- 0.35 }^{+ 0.33 }$ &   2.61 $_{- 0.25 }^{+ 0.23 }$ &   1.72 $\pm$ 0.20  &  1.07 $_{- 0.11 }^{+ 0.11 }$ \\ 
        3 & 5600 & 3.00 & G6 & 2.33 $_{- 0.31 }^{+ 0.28 }$ &   1.99 $_{- 0.24 }^{+ 0.21 }$ &   1.52 $\pm$ 0.20  &  0.78 $_{- 0.10 }^{+ 0.09 }$ \\ 
        4 & 5700 & 4.00 & G3 & 1.01 $_{- 0.29 }^{+ 0.26 }$ &   0.81 $_{- 0.23 }^{+ 0.20 }$ &   0.91 $\pm$ 0.20  &  0.35 $_{- 0.09 }^{+ 0.08 }$ \\ 
        5 & 5800 & 4.00 & G2 & 1.70 $_{- 0.28 }^{+ 0.25 }$ &   1.44 $_{- 0.22 }^{+ 0.19 }$ &   1.28 $\pm$ 0.20  &  0.56 $_{- 0.09 }^{+ 0.08 }$ \\ 
        6 & 5600 & 4.00 & G6 & 1.23 $_{- 0.30 }^{+ 0.26 }$ &   0.99 $_{- 0.24 }^{+ 0.21 }$ &   1.17 $\pm$ 0.20  &  0.48 $_{- 0.10 }^{+ 0.09 }$ \\ 
        7 & 5500 & 2.50 & G8 & 2.72 $_{- 0.36 }^{+ 0.29 }$ &   2.5 $_{- 0.25 }^{+ 0.22 }$ &   2.23 $\pm$ 0.20  &  0.88 $_{- 0.12 }^{+ 0.09 }$ \\ 
        8 & 5300 & 3.00 & K0 & 2.90 $_{- 0.33 }^{+ 0.32 }$ &   2.28 $_{- 0.27 }^{+ 0.24 }$ &   1.82 $\pm$ 0.20  &  0.90 $_{- 0.10 }^{+ 0.10 }$ \\ 
        9 & 5700 & 4.00 & G3 & 1.29 $_{- 0.29 }^{+ 0.26 }$ &   1.19 $_{- 0.23 }^{+ 0.20 }$ &   1.20 $\pm$ 0.20  &  0.43 $_{- 0.09 }^{+ 0.08 }$ \\ 
        10 & 5400 & 3.00 & G9 & 2.86 $_{- 0.34 }^{+ 0.29 }$ &   2.36 $_{- 0.26 }^{+ 0.22 }$ &   1.70 $\pm$ 0.20  &  0.87 $_{- 0.11 }^{+ 0.09 }$ \\ 
        11 & 5300 & 4.00 & K0 & 2.40 $_{- 0.32 }^{+ 0.30 }$ &   1.73 $_{- 0.27 }^{+ 0.24 }$ &   1.83 $\pm$ 0.20  &  0.66 $_{- 0.09 }^{+ 0.10 }$ \\ 
        12 & 5700 & 3.00 & G3 & 2.56 $_{- 0.29 }^{+ 0.29 }$ &   2.24 $_{- 0.22 }^{+ 0.21 }$ &   1.78 $\pm$ 0.20  &  0.77 $_{- 0.09 }^{+ 0.10 }$ \\ 
        13 & 6100 & 4.00 & F9 & 1.30 $_{- 0.25 }^{+ 0.25 }$ &   1.21 $_{- 0.19 }^{+ 0.18 }$ &   1.66 $\pm$ 0.20  &  0.40 $_{- 0.08 }^{+ 0.08 }$ \\ 
        14\tablenotemark{a} & 5800 & 4.00 & G3 & 2.07 $_{- 0.28 }^{+ 0.25 }$ &   1.54 $_{- 0.22 }^{+ 0.19 }$ &   1.52 $\pm$ 0.20  &  0.63 $_{- 0.09 }^{+ 0.08 }$ \\ 
        15\tablenotemark{a} & 5300 & 3.00 & K0 & 2.30 $_{- 0.33 }^{+ 0.32 }$ &   1.71 $_{- 0.27 }^{+ 0.24 }$ &   1.85 $\pm$ 0.20  &  0.69 $_{- 0.10 }^{+ 0.10 }$ \\ 
        16 & 5500 & 3.00 & G8 & 2.05 $_{- 0.32 }^{+ 0.29 }$ &   1.83 $_{- 0.24 }^{+ 0.22 }$ &   1.81 $\pm$ 0.20  &  0.52 $_{- 0.10 }^{+ 0.10 }$ \\ 
        17\tablenotemark{a} & 5500 & 3.00 & G8 & 2.31 $_{- 0.32 }^{+ 0.29 }$ &   1.90 $_{- 0.24 }^{+ 0.22 }$ &   1.09 $\pm$ 0.20  &  0.65 $_{- 0.10 }^{+ 0.10 }$ \\ 
        18 & 5600 & 4.00 & G6 & 1.39 $_{- 0.30 }^{+ 0.27 }$ &   1.09 $_{- 0.24 }^{+ 0.21 }$ &   1.39 $\pm$ 0.20  &  0.39 $_{- 0.10 }^{+ 0.09 }$ \\ 
        19 & 6100 & 3.00 & F9 & 2.70 $_{- 0.27 }^{+ 0.26 }$ &   2.22 $_{- 0.20 }^{+ 0.18 }$ &   1.12 $\pm$ 0.20  &  0.72 $_{- 0.09 }^{+ 0.08 }$ \\ 
        20 & 5000 & 3.00 & K2 & 1.42 $_{- 0.39 }^{+ 0.33 }$ &   1.41 $_{- 0.32 }^{+ 0.27 }$ &   2.54 $\pm$ 0.20  &  0.39 $_{- 0.12 }^{+ 0.10 }$ \\ 
        21 & 5400 & 3.00 & G9 & 2.38 $_{- 0.34 }^{+ 0.28 }$ &   1.92 $_{- 0.26 }^{+ 0.22 }$ &   1.38 $\pm$ 0.20  &  0.62 $_{- 0.11 }^{+ 0.09 }$ \\ 
        22 & 5400 & 3.00 & G9 & 2.77 $_{- 0.34 }^{+ 0.29 }$ &   1.95 $_{- 0.26 }^{+ 0.22 }$ &   1.60 $\pm$ 0.20  &  0.59 $_{- 0.11 }^{+ 0.09 }$ \\ 
        23 & 6100 & 4.00 & F9 & 1.64 $_{- 0.25 }^{+ 0.24 }$ &   1.37 $_{- 0.19 }^{+ 0.18 }$ &   1.26 $\pm$ 0.20  &  0.56 $_{- 0.08 }^{+ 0.08 }$ \\ 
        24 & 5700 & 4.00 & G3 & 0.79 $_{- 0.29 }^{+ 0.26 }$ &   0.69 $_{- 0.23 }^{+ 0.20 }$ &   0.55 $\pm$ 0.20  &  0.30 $_{- 0.09 }^{+ 0.08 }$ \\ 
        25 & 5800 & 4.00 & G3 & 1.14 $_{- 0.28 }^{+ 0.25 }$ &   1.16 $_{- 0.22 }^{+ 0.19 }$ &   1.15 $\pm$ 0.20  &  0.42 $_{- 0.09 }^{+ 0.08 }$ \\ 
        26 & 6100 & 4.00 & F9 & 0.96 $_{- 0.25 }^{+ 0.25 }$ &   1.04 $_{- 0.19 }^{+ 0.18 }$ &   1.38 $\pm$ 0.20  &  0.35 $_{- 0.08 }^{+ 0.08 }$ \\ 
        27 & 6100 & 4.00 & F9 & 0.84 $_{- 0.25 }^{+ 0.25 }$ &   0.87 $_{- 0.19 }^{+ 0.18 }$ &   0.99 $\pm$ 0.20 &  0.31 $_{- 0.08 }^{+ 0.08 }$ \\ 
        \hline
    \end{tabular}
\tablenotetext{a}{\textit{APASS} photometry was not available. \textit{Pan-STARRS} photometry was used instead with transformations from \citet{Jester2005}.}
\end{table}

\subsubsection{Comparisons Between Extinction Estimates with $R_{\rm{V}}$=3.1}
\label{ext_est_compare}
\begin{figure}[H]
    \centering
    \includegraphics[width=7in]{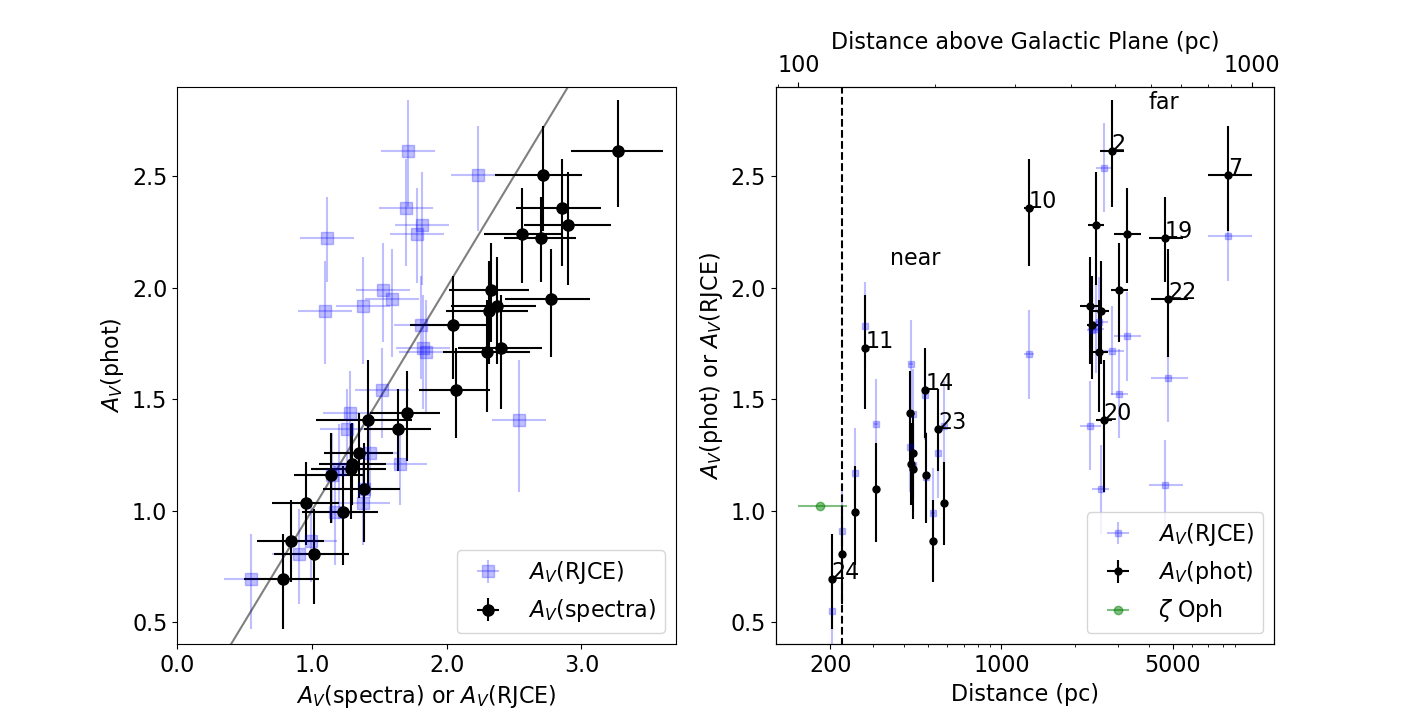}
    \caption{\textit{(left)} $A_{\rm{V}}$(phot) versus $A_{\rm{V}}$(spectra) in black circles and $A_{\rm{V}}$(phot) versus $A_{\rm{V}}$(RJCE) in blue squares. The solid line shows what would be a 1:1 correspondence between estimates. \textit{(right)} $A_{\rm{V}}$(phot) in black and  $A_{\rm{V}}$(RJCE) in blue versus distance and distance above the Galactic Plane. Some target numbers are shown for clarity. The green point marks $\zeta$~Oph, and the near and far groups are labeled.  $A_{\rm{V}}$(RJCE) is consistently smaller than $A_{\rm{V}}$(phot) at large extinctions and distances.} 
    \label{Av}
\end{figure}

To examine the validity of each $A_{\rm{V}}$ estimate, we compared the three values for each target in Figure~\ref{Av}. The left panel plots $A_{\rm{V}}$(phot) versus $A_{\rm{V}}$(spectra) in black circles and $A_{\rm{V}}$(phot) versus $A_{\rm{V}}$(RJCE) in blue squares. The line shows the 1:1 correspondence between the estimates. The black points show that $A_{\rm{V}}$(spectra) and $A_{\rm{V}}$(phot) are tightly correlated with a slope slightly less than unity. The strong correlation, with a Pearson correlation coefficient of 0.97, between $A_{\rm{V}}$(spectra) and $A_{\rm{V}}$(phot) affirms the consistency of these independent optically-based measurements. The blue points show that $A_{\rm{V}}$(phot) and  $A_{\rm{V}}$(RJCE) are correlated at small extinctions but then diverge, falling consistently above the 1:1 relation at higher extinctions. $A_{\rm{V}}$(phot) and  $A_{\rm{V}}$(RJCE) are not well-correlated, with a Pearson correlation coefficient of 0.57 for all target stars.


The right panel of Figure~\ref{Av} shows $A_{\rm{V}}$(phot) and $A_{\rm{V}}$(RJCE) versus distance (lower x-axis) and distance above the plane (upper x-axis). Black points show $A_{\rm{V}}$(phot) and blue points show $A_{\rm{V}}$(RJCE). The green point designates $\zeta$~Oph ($A_{\rm{V}}$=1.02 mag, $d$=182 pc), with its extinction arising within multiple foreground cloud complexes and an \ion{H}{2} region \citep{Wood_2005, choi}. The dashed vertical line marks the scale height of dust in the Galactic Plane, $z = 125^{+7}_{-7}$~pc \citep{Marshall_2006}.  In the near group of targets, $A_{\rm{V}}$(RJCE) is similar to  $A_{\rm{V}}$(phot), with means of 1.27$\pm 0.09$ mag and 1.17$\pm 0.08$ mag, respectively (uncertainties represent the error of the mean). In the far group, $A_{\rm{V}}$(RJCE) consistently falls below $A_{\rm{V}}$(phot) with means of 1.70$\pm 0.11$ mag and 2.05$\pm 0.09$ mag, respectively. Figure \ref{Av} indicates that $A_{\rm{V}}$(RJCE) is systematically less than $A_{\rm{V}}$(phot) in the far group. 

Regardless of measure used, it is clear that the extinction increases with distance. The far group has an average $A_{\rm{V}}$(phot) 0.88~mag larger than the near group and an average $A_{\rm{V}}$(RJCE) 0.43~mag larger than the near group. The Pearson correlation coefficient between distance and $A_{\rm{V}}$(phot) for all targets is 0.73, implying positive linear correlation.  The correlation between distance and $A_{\rm{V}}$(RJCE) is smaller: 0.5 for all targets. By examining the near and far group separately, we find that correlation in the far group is lower than in the near group for both $A_{\rm{V}}$(phot) and $A_{\rm{V}}$(RJCE). The difference between $A_{\rm{V}}$(phot) and $A_{\rm{V}}$(RJCE) in the far group suggests that the extinction depends on the wavelength range used for its calculation (visible vs. IR photometry), which indicates a departure from the average reddening curve in the Milky Way.

\subsubsection{Determining $E(B-V)$ and Incrementally Dereddening Along the Line of Sight}

The discrepancy between extinction estimates (primarily in the far group) motivates a technique to explore the properties of the different dust populations with distance. We therefore developed a tool to perform Incremental DeReddening Along the Line of Sight (IDEALS), allowing us to investigate both $A_{\rm{V}}$ and $R_{\rm{V}}$ as a function of distance.

Following \citet{Fitzpatrick2019}, we constructed the extinction curve for the targets over optical and infrared wavelengths. Using the B, V, and g$^{\prime}$r$^{\prime}$i$^{\prime}$z$^{\prime}$y$^{\prime}$ magnitudes from the \textit{APASS} database (or \textit{Pan-STARRS} when required) and the 2MASS JHK and \textit{Spitzer} 3.6~$\mu$m and 4.5~$\mu$m magnitudes, we normalized the stellar model from Table \ref{exttab} and photometry at V band and performed synthetic photometry on each model as described for $A_{\rm{V}}$(phot). For comparison to the previous estimates, we assumed the \citet{Fitzpatrick2019} empirical extinction curves and fit for the best $A_{\rm{V}}$; however, we also allowed $R_{\rm{V}}$ to vary. We directly calculated the color excess, $E(\lambda-V)$, at each bandpass relative to V band in order to encompass both $A_{\rm{V}}$ and $R_{\rm{V}}$ into one measure. Table \ref{exttab} column 7 lists $E(B-V)$ for each target. We then expressed the reddening relative to V band in the standard nomenclature from \citet{Johnson1965},
\begin{equation}
    k(\lambda-V)~= \frac{E(\lambda-V)}{E(B-V)}.
\end{equation}

\begin{figure}
    \centering
    \includegraphics[width=6in]{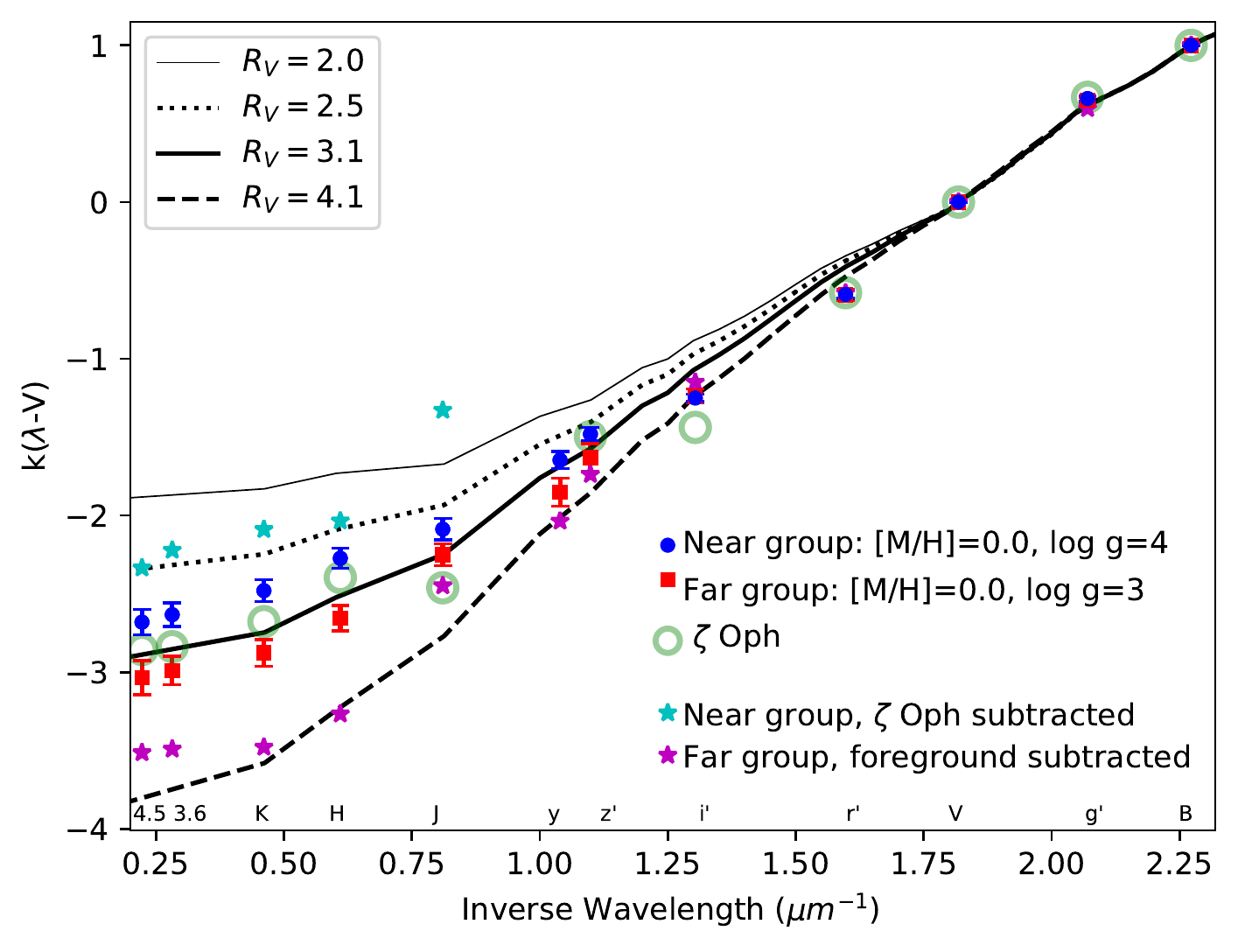}
    \caption{Extinction curves represented by the reddening relative to V-band, i.e., E($\lambda$-V) per $E(B-V)$ versus inverse wavelength in microns.  Blue points mark the mean and error of the mean for the near group of 14 stars, red points mark the mean of the far group of 13 stars, cyan stars are the mean of the near group with the reddening of $\zeta$~Oph subtracted, and magenta stars are the mean of the far group with the reddening of the near group subtracted. Green circles represent $\zeta$~Oph. Tracks denote the canonical Milky Way reddening curves from \citet{Fitzpatrick2019} for a range of $R_{\rm V}$=2.0/2.5/3.1/4.1, according to the legend. }
    \label{fig:klamplot}
\end{figure}

Figure \ref{fig:klamplot} plots $k(\lambda-V)$, the extinction curve, versus inverse wavelength. Black lines show the extinction curves from \citet{Fitzpatrick2019}\footnote{These closely match the \citet{CCM89} analytic extinction curves.}. Line styles denote different $R_{\rm{V}}$, with larger $R_{\rm{V}}$ creating a steeper slope in $k(\lambda-V)$. Labels along the bottom mark standard bandpasses at their corresponding inverse wavelength. Blue and red points denote average values of $k(\lambda-V)$ for the near and far groups, respectively. Error bars represent the error of the mean within each band. Any systematic errors on our adopted stellar $T_{\rm eff}$ would merely shift the points up or down together relative to the reddening curves. While they are similar in the optical, the near group of targets (blue) follows a smaller $R_{\rm{V}}$ trend than the far group (red) in the infrared. However, both groups are close to the Milky Way average of $R_{\rm{V}}$=3.1, shown by the solid black line.

The green circles in Figure \ref{fig:klamplot} show the extinction curve of $\zeta$~Oph. Owing to its brightness, there are no measurements of $\zeta$~Oph on a modern photometric system. Accordingly, we adopted optical BVRI magnitudes from \citet{Johnson1965} and used the equations of \citet{Jester2005} and \citet{Fernie1983} to transform these into  g$^{\prime}$r$^{\prime}$i$^{\prime}$z$^{\prime}$ magnitudes.  In the infrared we used the 2MASS JHK and {\it Spitzer} 3.6 and 4.5~$\mu$m fluxes from \textit{IRSA} and employed the corresponding zero points to obtain magnitudes.  At 3.6 and 4.5~$\mu$m the {\it Spitzer} fluxes are $\sim$30\% smaller than the ground-based measurements of \citet{Gehrz1974}, so we adopted the average of the two fluxes.  For the intrinsic spectral energy distribution we selected the $T_{\rm eff}$~=~31,000~K, $\log$~$g$~=~4.5 ``BT-settl'' model atmosphere of \citet{Allard2011} which yields very similar results to the older \citet{Castelli2003} models. Despite larger uncertainties resulting from the photometric transformations, the reddening curve for $\zeta$~Oph is consistent with mean interstellar $R_{\rm V}$=3.1 dust, indicating that the dust in the $d$$<$182~pc foreground is ``average''..   

To examine $R_{\rm{V}}$ with distance, we constructed reddening curves for the near and far groups with their average foreground reddening removed. This is the crux of the IDEALS technique: it can explore dust properties at different distances along a line of sight by incrementally removing a well-determined foreground. Cyan stars in Figure \ref{fig:klamplot} show the near group with the reddening of $\zeta$~Oph subtracted. With the foreground ($\zeta$~Oph) removed, the near group follows a smaller mean $R_{\rm{V}}$$\approx$2.4. Magenta stars show the far group with the average reddening of the near group subtracted. After subtracting the average of the near group, the far group has a noticeably greater mean $R_{\rm{V}}$$\approx$3.6.

\subsection{A Highly Polarizing Dust Component}
\begin{figure}
    \centering
    \includegraphics[width=4in]{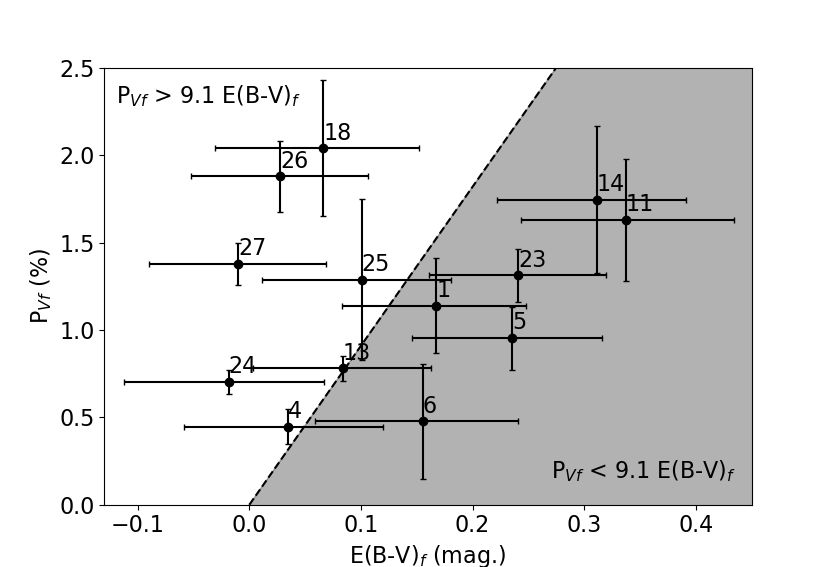}
    \caption{Percent polarization versus color excess with the foreground of $\zeta$~Oph removed from each. The dashed line shows the observed maximum polarization per unit reddening of dust, $P_{\rm{V}}$=9.1$E(B-V)$ \citep{serkowski}. Many targets in the nearby group fall above the line and are therefore highly polarized for the amount of reddening.}
    \label{fig:pol-ext}
\end{figure}

Observations of 180 sightlines from \citet{serkowski} indicate that while polarization and color excess are not correlated, there is a maximum amount of polarization per unit reddening of $P_{\rm{V}}$=9.1$E(B-V)$. In their sample, only rarely do stars fall above this limit. More recent measurements \citep[for example, ][]{Andersson} continue to support this observational upper limit. Figure \ref{fig:pol-ext} plots $P_{\rm{V f}}$ versus $E(B-V)_{\rm{f}}$ (the foreground-subtracted color excess, adopting $E(B-V)$=0.33 mag. for $\zeta$ Oph) for the near group to examine the polarization efficiency of dust between $\zeta$~Oph and the stars in the immediate background. The dashed line shows the observational limit proposed by \citet{serkowski}, and the shaded region denotes the ``permitted'' region below this limit. The foreground-subtracted polarizations (labeled by target number) in the near group are high, often above the limit. This suggests that the near group of stars lies behind or is embedded within a population of dust grains more efficiently polarizing than the typical ISM.

The negative color excess of target \#24, along with its close distance ($d=203$~pc),  may indicate that it lies in the foreground of $\zeta$~Oph, which is entirely possible when taking into account the uncertainties in the $Gaia$ distance of  $\zeta$~Oph. To analyze this possibility, we replicated Figure \ref{fig:pol-ext} with the color excess and polarization of target \#24 as the values for foreground subtraction. Target \#24 has an $E(B-V)$=$0.30^{+0.08}_{-0.09}$ and a raw polarization of $P_{\rm{V}}$=1.09$\pm$0.07\% at 111.9$\pm$1.82\degr. Although the group of targets in Figure \ref{fig:pol-ext} shifts slightly to the right, the figure and foreground-subtracted polarizations and color excesses remain essentially the same despite the choice of foreground reference star. 

\subsection{The Bowshock and Striation Dust Structures}
\label{section_striations}
While the bowshock is certainly in close proximity to and associated with $\zeta$~Oph, the location and properties of the infrared striations present in Figures \ref{rawpol} and \ref{subpol} are less clear. Both the bowshock and striations appear in all three wavelengths of the \textit{Spitzer} images, but most strongly at 8~$\mu$m, a wavelength dominated by Polycyclic Aromatic Hydrocarbons (PAHs) that are excited by the UV radiation from hot stars. The clear signature of PAH emission from the striations demands illumination by UV photons, circumstantially placing them in close proximity to $\zeta$~Oph. As an O9.2IV star ($T_{\rm eff}$=31,000~K, $R_\star$=7.2~$R_\odot$), $\zeta$~Oph dominates the radiant energy in its vicinity. From the catalog of Galactic OB stars \citep{Reed2003}, we find no other O or early B stars within 50~pc of $\zeta$~Oph. At a distance of $z$$\sim74$~pc above the Plane, $\zeta$~Oph is the leading candidate for the source of far-UV photons exciting the PAH molecules \citep{Tielens2008} in the striations. The ratio of specific intensities at 8 and 24~$\mu$m, $I_{\rm 8}$/$I_{\rm 24}$, is $\approx$0.14 for the bowshock (redder in Figure~\ref{subpol}) and $\approx$0.20 for the striations (greener in Figure~\ref{subpol}), suggesting a larger PAH contribution in the latter. Likewise, the 8~$\mu$m to 4.5~$\mu$m (a band free of PAH contributions) ratio increases from $\approx$8.5 in the bowshock to $\approx$20 in the striations. This is similar to the value observed in established high-PAH photon-dominated regions \citep{Churchwell2006,Deharveng2009}.  The disparate colors between the bowshock and striations indicate two distinct dust structures having a different set of physical conditions and possibly different grain properties.

To quantify the orientation of the dust striations and compare them to the polarization position angles in this field, we employed the Rolling Hough Transform from \citet{Clark_2014ApJ...789...82C}. Their code reads images and calculates $R(\theta, x, y)$, a function that encodes intensity as a function of angle, $\theta$, for components in the image at locations $x$ and $y$. We performed this process using regions of the 8~$\mu$m \textit{Spitzer} image lying directly above and below the prominent bowshock structure, excluding the bowshock because of its high intensity and dominance over the diffuse striations. By integrating $R(\theta, x, y)$ for each $x$ and $y$ with Equations 7 and 8 from \citet{Clark_2014ApJ...789...82C},
we obtained the overall orientation angles for the striated regions directly above and below the bowshock. In Figure \ref{subpol}, the blue dashed lines show the average orientation of the striations above and below the bowshock, 51.6\degr\ and 57.2\degr, respectively.

\section{Discussion}
\label{section_discuss}
The extinction and polarization data provide a clear picture of the dust along these lines of sight. While the foreground dust masks most of the variation in the form of the extinction curve, removal of that component reveals the interesting structures beyond. 

\subsection{Dust Populations Along the Line of Sight}

The correlation between $A_{\rm{V}}$(RJCE) and  $A_{\rm{V}}$(phot) decreases at greater extinctions and distances. This is clearly demonstrated in Figure \ref{Av} and in the correlation coefficients calculated in Section \ref{ext_est_compare}. This indicates---most dramatically at larger extinctions--- {\it a departure of the reddening curve from the Milky Way average, the shape of which is assumed in the three methods} used in Figure~\ref{Av}. We conclude from Figure~\ref{Av} that the reddening curve for the high-extinction, far group differs from that of the low-extinction near group.

The $k(\lambda-V)$ trends in Figure \ref{fig:klamplot} suggest that the near group of stars is reddened by a population of dust with $R_{\rm{V}}$$\approx$2.4, while the far group is reddened by a population of dust with $R_{\rm{V}}$$\approx$3.6. The small $R_{\rm{V}}$ in the near group indicates an excess of small grains (as discussed in Section \ref{section_intro}) just behind $\zeta$~Oph in a ``mid-distance cloud''. An excess of large grains (or deficit of small grains) above the Galactic Plane, located in a ``distant cloud'', could plausibly explain the high $R_{\rm{V}}$ in the far group. On its own, the far group shows minimal correlation between extinction and distance (Pearson correlation coefficient of 0.47 for $A_{\rm{V}}$(phot) and 0.21 for $A_{\rm{V}}$(RJCE)) in Figure~\ref{Av}. This indicates that the additional extinction affecting the far group lies in the 600--2000~pc gap separating the two groups, thereby extincting the far group somewhat uniformly, although the dispersion of extinctions suggests some patchiness in dust coverage. Of course, both of these dust components idealized as discrete slabs may actually consist of smaller  constituent clouds.

\begin{figure}
    \centering
    \includegraphics[width=7.5in]{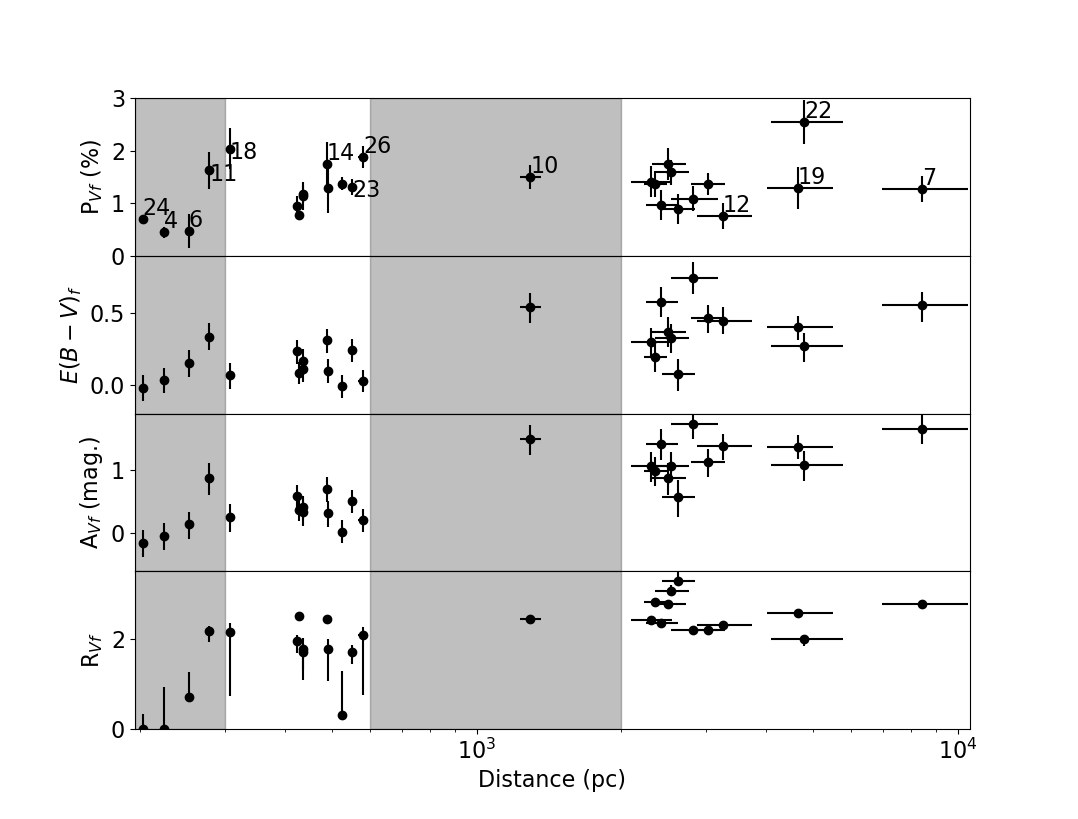}
    \caption{$P_{\rm{V f}}$, $E(B-V)_{\rm{f}}$, $A_{\rm{V f}}$, and $R_{\rm{V f}}$ with distance. Gray regions represent likely locations of dust along the line of sight, the highly polarizing mid-distance dust and the non-polarizing dust at a greater distance. Only some target numbers are shown for clarity.}
    \label{fig:fitz_dist}
\end{figure}

Figure \ref{fig:fitz_dist} illustrates the preferred configuration of dust along the line of sight. The top panel shows $P_{\rm{V f}}$ versus distance. The second panel shows $E(B-V)_{\rm{f}}$, the foreground-subtracted color excess. The two bottom panels show $A_{\rm{V f}}$ and $R_{\rm{V f}}$, which were obtained by dereddening using the foreground parameters of $\zeta$~Oph\footnote{$E(B-V)$=0.33 from \citet{Ducati_2002yCat.2237....0D, Pecaut_2013}} and refitting the \citet{Fitzpatrick2019} extinction curve. Shaded gray regions represent the proposed two discrete dust populations that lie in the background of $\zeta$ Oph: a highly polarizing mid-distance 200--300 pc component and a non-polarizing 600--2000 pc distant component.

In the top panel, $P_{\rm{V f}}$ increases sharply just beyond targets \#4, \#6, \#24 (203--252~pc) from 0.5\% to as high as 2\%. The substantial increase in polarization occurs in the same dust that creates only a small increase in color excess ($\Delta$E(B-V)$\approx$0.08, visible in the second panel). This suggests a mid-distance dust component with high grain alignment efficiency somewhere in the vicinity of 200--300~pc. {\it These sightlines raise the interesting possibility that the bulk of interstellar polarization may take place within a very small volume fraction of the ISM if such highly polarizing dust structures are common.}  The polarization then levels off below 2\% at large distances. Although there is some dispersion in the far group, the polarization is consistently between 1--2\%.    

As distance increases, $E(B-V)_{\rm{f}}$ also increases for the first few targets, then flattens for the remainder of the near group. The dispersion in $E(B-V)_{\rm{f}}$ becomes greater at larger distances, and the mean $E(B-V)_{\rm{f}}$ is larger than in the near group. $A_{\rm{V f}}$ reflects the same trends as $E(B-V)_{\rm{f}}$, increasing consistently with distance. Because $E(B-V)_{\rm{f}}$ and $A_{\rm{V f}}$ increase with distance, but $P_{\rm{V f}}$ does not, we conclude that the distant dust population (located between 600-2000~pc) extincts background light but does not create a noticeable polarization signature. This lack of polarization signature at large distances may stem from a tangled magnetic field, a weak or absent magnetic field, or a magnetic field oriented along the line of sight. We are not able to distinguish which of these scenarios is at work, so we simply refer to this dust component as non-polarizing. 

The bottom panel shows $R_{\rm{V f}}$, the best fit for $R_{\rm{V}}$ after dereddening by the foreground of $\zeta$~Oph. For the targets with the lowest $A_{\rm{V f}}$ and nearest distances (\#4 and \#6), fits of $R_{\rm{V f}}$ are unreliable because there is very little reddening to fit, so $R_{\rm{V f}}$ is set arbitrarily to zero. The $R_{\rm{V f}}$ of the near group is low (1.8$\pm$0.7), matching the $R_{\rm{V}}$$\approx$2.4 obtained for the foreground-subtracted near group in Figure \ref{fig:klamplot}. $R_{\rm{V f}}$ increases with distance towards the far group to 2.6$\pm$0.4. After further dereddening by the average extinction of the near group, the far group shifts towards a larger $R_{\rm{V}}$$=$3.1$\pm$1.0 (matching the $R_{\rm{V}}$$\approx$3.6 obtained in Figure \ref{fig:klamplot}).

While an average $R_{\rm{V}}$ for the Milky Way is established, it is based on O-, B-, and A-type stars near the mid-Plane and does not account for possible variations throughout the Galaxy \citep{mw_rv, Sneden_1978}. $R_{\rm{V}}$ is known to vary between values near two \citep{Larson_2000ApJ...532.1021L} to more than six \citep{Johnson1965}. \citet{Whittet} discovered variations with Galactic longitude that suggest changes in the size distribution of dust grains in the local spiral arm.  Dust at high Galactic latitudes is known to be inconsistent with standard reddening laws, and no single $R_{\rm{V}}$ can explain observations of both the optical and ultraviolet extinctions toward extragalactic sources \citep{Peek_2013}. It is fair to assume that dust properties vary throughout the Milky Way, possibly changing significantly along sightlines outside of the Galactic Plane like $\zeta$~Oph ($b$=24\degr). HD~210121 (B7II) lies below the Galactic Plane and is also obscured by a diffuse cloud. The sightline towards that star has a low $R_{\rm{V}}$$\approx$2.1 \citep{Larson_2000ApJ...532.1021L}, demonstrating that the smaller-than-average $R_{\rm{V}}$ in the near group may be a more global trend for diffuse clouds like those towards $\zeta$~Oph. In fact, many high latitude clouds have enhanced abundances of relatively small dust grains \citep{Larson_2005ApJ...623..897L}.

\subsection{Magnetic Field Characteristics and Relation to Dust Structures}


Figure~\ref{fig:pol-ext} indicates that the mid-distance dust directly behind $\zeta$~Oph is highly polarizing. It is likely that the high polarization per reddening arises from dust behind $\zeta$~Oph, as that star itself falls far below the Serkowski limit, with $E(B-V)$=0.33 and $P_{\rm{V}}$=1.44\% \citep{serkowski}. This highly polarizing component is not entirely novel, as \citet{Gontcharov_2019MNRAS.483..299G} found sources in an all-sky polarization survey with $P_{\rm{V}}>$9.1$E(B-V)$. They explain that this may be the result of an underestimate in  $E(B-V)$, a non-standard reddening law, or simply a violation of the observational upper limit. In our case, the nearby group of stars has an $E(B-V)_{\rm f}$ that is well-constrained. The high $P_{\rm{V}}/E(B-V)$ ratio, which is substantially in excess of the limit from \citet{serkowski}, may then be the result of the non-standard reddening law seen in the near group or dust that is more efficiently polarized than typical ISM material.

The polarization vectors in Figure \ref{subpol} show the direction of the projected magnetic field with an average P.A.=77\degr\ and a dispersion of 18\degr. The Rolling Hough Transform provides an average direction for the striations below and above the bowshock (blue dashed lines in Figure \ref{subpol}) of 57.2\degr\ and 51.6\degr, respectively. This indicates that the polarization vectors are aligned with the infrared dust striations in the field of view. Taken together, the infrared images, polarization data, and striation orientation reveal a striated (P.A.$\approx$55\degr) PAH-emitting cloud spanning the field of view in close proximity to $\zeta$~Oph that hosts the aligned grains that polarize background starlight at a similar position angle.

The main arc of the bowshock is approximately perpendicular to the projected magnetic field, while the proper motion of $\zeta$ Oph is parallel. The magnetic field does not appear to follow the filaments in the bowshock, nor is there any apparent deviation in P.A. even near the highest surface brightness portions of the nebula.  The location and morphology of the bowshock are likely a result of the motion of $\zeta$~Oph and do not depend on the magnetic field orientation. Evidently, the bowshock contributes negligibly to the polarization, while the background striated medium is the strongly polarizing agent.

\subsection{An Optically Thin Bowshock Nebula}
A brief examination of specific targets shows that there are small-scale angular variations across the field of view, as well as variations along adjacent lines of sight. For example, Target \#22 has a significantly larger $P_{\rm{V f}}$ than target \#19 (2.56\% versus 1.25\%), despite the smaller $E(B-V)_{\rm{f}}$ (0.27 versus 0.40), similar location (within $\approx$2\arcmin), similar P.A.$_{\rm{f}}$ (64\degr\ versus 76\degr), and almost identical distance (4790~pc versus 4645~pc). Targets \#11 and \#12 lie within 30\arcsec\ of each other, but at very different distances (278~pc and 3246~pc, respectively). Target \#12 is understandably more reddened at a large distance, but has a smaller $P_{\rm{V f}}$ (0.45\% versus 1.60\% for \#11) and very different PA$_{\rm{f}}$ (110\degr\ versus 83\degr\ for \#11) than surrounding targets. The areal density of background sources is not sufficient to probe small-scale dust variations in a rigorous manner, but the variations on small angular scales and with distance are clear.

We have not sampled enough targets to determine if the bowshock nebula affects extinction. We find no clear difference in extinction between targets located directly behind (on) the bowshock versus off the bowshock. It is likely that distance is the primary parameter driving extinction, as seen in Figure \ref{fig:fitz_dist}. However, given that the optical depth of the bowshock at optical and UV wavelengths is expected to be small, it is not surprising that the prominent infrared nebulae does not produce an obvious signature of increased extinction. The median ratio of nebular far-infrared luminosity to stellar luminosity is $L_{\rm IR}$/L$_*\approx$0.003 in bowshock nebulae ($<$0.001 in the case of $\zeta$~Oph), indicating that only a small fraction of the stellar energy is intercepted and re-radiated by the dust in the nebula \citep{Kobulnicky_2017}.  The vast majority of infrared bowshock nebulae are optically thin, at least globally, a conclusion echoed by \citet{Henney_2019}.  This does not rule out the possible existence of optically thick clumps on small scales, but it does constrain them to have a small covering factor. The probability that one of the observed sightlines toward background stars would intercept such a clump is small, so the lack of an unambiguous signature in the extinction or polarization of background targets is unsurprising.

\section{Summary}
\label{section_summary}

Using optical polarimetry and spectrophotometry, we have examined the polarization and reddening along the lines of sight to the 27 background stars surrounding the O9.2IV star $\zeta$~Oph. We have developed a technique to incrementally subtract foreground reddening and polarization. This has allowed us to characterize the dust along the line of sight to the target stars, probing differences that would otherwise be masked by a sightline average. 

\begin{enumerate}
\item The dust characteristics along the sightlines surrounding $\zeta$~Oph vary with distance and can be represented as three distinct populations: the foreground dust ($R_{\rm{V}}$$\approx$3.1), the highly polarizing mid-distance dust ($R_{\rm{V}}$$<$3.1), and the non-polarizing distant dust ($R_{\rm{V}}$$>$3.1). The superposition of different dust populations along the line of sight leads to a sightline-averaged $R_{\rm{V}}$ that is close to typical for the ISM, masking pronounced variations in the reddening curve with distance. 

\item After subtracting the foreground characterized by the polarization of $\zeta$~Oph, the position angles become approximately parallel to the PAH-emitting infrared striations. The polarization most likely arises from the mid-distance dust responsible for those striations. We conclude that the dust is located behind---but in close proximity---to $\zeta$~Oph and is illuminated by the star's UV irradiation. With the foreground $R_{\rm{V}}$=3.1 dust removed, the mid-distance dust has an $R_{\rm{V}}$$\approx$2.4, suggesting an overabundance of small dust grains. 

This mid-distance dust (200--300~pc) is highly polarizing. It has a polarization efficiency larger than the limit observed by \citet[][$P_{\rm{V}}$/$E(B-V)$$>$9.1]{serkowski}. The high $P_{\rm{V f}}$/$E(B-V)_{\rm{f}}$ toward a number of stars over a limited path length ($\approx$200~pc) indicates that large changes in polarization can occur over small increments in dust column density. This suggests that a very small fractional ISM volume containing populations of unusually small, highly aligned grains may dominate the interstellar polarization signature.    

\item The distant dust (600--2000~pc) does not create a distinct polarization signature and has $R_{\rm{V}}$$\approx$3.6. The lack of polarization may stem from many factors, like a non-existent or tangled magnetic field or a magnetic field oriented along the line of sight. The larger-than-average $R_{\rm{V}}$ implies an excess of large dust grains at large distances above the Galactic Plane ($z$$>$250~pc). In the simplest case, this may hold true for the entire Milky Way above the Plane. 

\end{enumerate}

A complete understanding of variations in dust properties is essential  in correcting for extinction and motivates further investigation into $R_{\rm{V}}$ above the Plane to obtain a clear picture of dust properties throughout the Galaxy. The reddening map from \citet{Green_2019} is dependent on $R_{\rm{V}}$, and they emphasize that extinction maps derived from their reddenings without knowledge of the spatial dependence of $R_{\rm{V}}$ will introduce large-scale, spatially dependent systematic errors. Precision supernova surveys \citep{Betoule_2014, Scolnic_2018} and follow-up of optical transients from neutron star and black hole mergers \citep{Abbott_2017, Anand_2021} are just two examples of programs that require accurate corrections for extinction. The onslaught of accurate distances from $Gaia$ and publicly available spectral and photometric data make future studies of the extra-planar extinction curve practical, while these applications make it pressing.

\acknowledgments
We thank Matt Povich and Dan Clemens for helpful exchanges that contributed to this manuscript.
This research has made use of the NASA/IPAC Infrared Science Archive, which is funded by the National Aeronautics and Space Administration and operated by the California Institute of Technology. Ashley N. Piccone acknowledges support from NASA through grant 80NSSC21K1847.

\facilities{WIRO, APO, ARC, Spitzer, IRSA}
\software{IRAF \citep{Tody_1986SPIE..627..733T}, LACosmics \citep{Vandokkum}, PyHammer \citep{pyhammer}, Rolling Hough Transform \citep{Clark_2014ApJ...789...82C}}

\bibliography{mybib}

\end{document}